\documentclass[final,5p,times]{elsarticle}

\usepackage{graphicx}
\usepackage{GRAF}

\newcommand{\degree}{\ensuremath{^\circ}}
\newcommand{\CesrTA}{\textsc{CesrTA}}

\def\gappeq{\mathrel{ \rlap{\raise.5ex\hbox{$>$}}
                      {\lower.5ex\hbox{$\sim$}}  } }
\def\lappeq{\mathrel{ \rlap{\raise.5ex\hbox{$<$}}
                      {\lower.5ex\hbox{$\sim$}}  } }

\makeatletter
\def\appendixname{Appendix}
\renewcommand{\appendix}{\par
  \setcounter{section}{0}%
  \setcounter{subsection}{0}%
  \setcounter{equation}{0}%
  \setcounter{figure}{0}%
  \setcounter{table}{0}%
  \gdef\thefigure{\@Alph\c@section.\arabic{figure}}%
  \gdef\thetable{\@Alph\c@section.\arabic{table}}%
  \gdef\thesection{\appendixname~\@Alph\c@section}%
  \@addtoreset{equation}{section}%
  \@addtoreset{figure}{section}%
  \@addtoreset{table}{section}%
  \gdef\theequation{\@Alph\c@section.\arabic{equation}}%
  \addtocontents{toc}{\string\let\string\numberline\string\tmptocnumberline}{}{}
}
\long\def\MaketitleBox{%
  \resetTitleCounters
  \def\baselinestretch{1}%
  \begin{center}%
   \def\baselinestretch{1}%
    \Large\@title\par\vskip18pt
    \normalsize\elsauthors\par\vskip10pt
    \footnotesize\itshape\elsaddress\par\vskip18pt
    \hrule\vskip12pt
    \ifvoid\absbox\else\unvbox\absbox\par\vskip10pt\fi
    \ifvoid\keybox\else\unvbox\keybox\par\vskip10pt\fi
    \hrule\vskip12pt
    \end{center}%
  }
  \providecommand\bibsection{}
  \renewcommand\bibsection{%
   \section*{\refname\@mkboth{\MakeUppercase{\refname}}{\MakeUppercase{\refname}}}%
   \addcontentsline{toc}{section}{\refname}%
  }%
\makeatother
\renewcommand{\today}{\number\day\space \ifcase\month\or
  January\or February\or March\or April\or May\or June\or
  July\or August\or September\or October\or November\or December\fi
  \space \number\year}

\biboptions{sort&compress}

\journal{Nucl. Instrum. Methods Phys. Res. A}

\usepackage{ifthen}

\RequirePackage[colorlinks=true,hyperfootnotes=false,bookmarksnumbered=true]{hyperref}
\RequirePackage[all]{hypcap}
\RequirePackage[capitalise,nameinlink]{cleveref}

\hypersetup{pdftitle={In-Situ Measurements of SEY in an Accelerator Environment: Instrumentation \& Methods},
            pdfauthor={W. Hartung et al.},
            pdfkeywords={secondary emission, electron cloud, beam scrubbing, storage ring, damping ring}}
\begin{document}

\begin{frontmatter}


\title{In-Situ Measurements of the Secondary Electron Yield in an
  Accelerator Environment: Instrumentation and Methods}


\author{W.~H.~Hartung}
\author{D.~M.~Asner\fnref{pnnl}}
\fntext[pnnl]{Present address: Pacific Northwest National Laboratory, Richland, WA}
\author{J.~V.~Conway}
\author{C.~A.~Dennett\fnref{mit}}
\fntext[mit]{Present address: Department of Nuclear Science and Engineering, Massachusetts Institute of Technology, Cambridge, MA}
\author{S.~Greenwald}
\author{J.-S.~Kim\fnref{prince}}
\fntext[prince]{Present address: Department of Electrical Engineering, Princeton University, Princeton, NJ}
\author{Y.~Li}
\author{T.~P.~Moore}
\author{V.~Omanovic}
\author{M.~A.~Palmer\fnref{fnal}}
\fntext[fnal]{Present address: Fermi National Accelerator Laboratory, Batavia, IL}
\author{C.~R.~Strohman}

\address{Cornell Laboratory for Accelerator-based ScienceS and
  Education, Cornell University, Ithaca, New York, USA}

\begin{abstract}
The performance of a particle accelerator can be limited by the
build-up of an electron cloud (EC) in the vacuum chamber.  Secondary
electron emission from the chamber walls can contribute to EC growth.
An apparatus for in-situ measurements of the secondary electron yield
(SEY) in the Cornell Electron Storage Ring (CESR) was developed in
connection with EC studies for the CESR Test Accelerator program.  The
CESR in-situ system, in operation since 2010, allows for SEY
measurements as a function of incident electron energy and angle on
samples that are exposed to the accelerator environment, typically
5.3~GeV counter-rotating beams of electrons and positrons.  The system
was designed for periodic measurements to observe beam conditioning of
the SEY with discrimination between exposure to direct photons from
synchrotron radiation versus scattered photons and cloud electrons.
The samples can be exchanged without venting the CESR vacuum chamber.
Measurements have been done on metal surfaces and EC-mitigation
coatings.  The in-situ SEY apparatus and improvements to the
measurement tools and techniques are described.
\end{abstract}


\end{frontmatter}

%
\newlength{\narrowwidth}%
\newlength{\middlewidth}%
\newlength{\widewidth}%
\newlength{\widestwidth}%
\newlength{\heightone}%
\setlength{\heightone}{0.3\textheight}%
\newlength{\heighttwo}%
\setlength{\heighttwo}{0.35\textheight}%
\ifthenelse{\lengthtest{\columnwidth=\textwidth}}
{
\setlength{\narrowwidth}{0.6\textwidth}
\setlength{\middlewidth}{0.7\textwidth}
\setlength{\widewidth}{0.85\textwidth}
\setlength{\widestwidth}{\textwidth}
\setlength{\heightone}{0.8\heightone}
\setlength{\heighttwo}{0.8\heighttwo}}
{
\setlength{\narrowwidth}{\columnwidth}
\setlength{\middlewidth}{\columnwidth}
\setlength{\widestwidth}{\columnwidth}
\setlength{\widewidth}{\columnwidth}}
%
\newlength{\smallspace}%
\newlength{\medspace}%
\newlength{\bigspace}%
\newlength{\biggerspace}%
\setlength{\smallspace}{0.5ex}
\setlength{\medspace}{2ex}
\setlength{\bigspace}{3.5ex}
\setlength{\biggerspace}{5ex}


\section{Introduction\label{S:intro}}

Ideally, the beams in a particle accelerator propagate through a
perfectly evacuated chamber.  In reality, the vacuum chamber contains
residual gas, ions, and low-energy electrons.  Low-energy electrons
can be produced by photo-emission when synchrotron radiation photons
strike the wall of the chamber; by bombardment of the wall by the beam
halo; or by ionization of residual gas by the beam.  If the electrons
hit the wall and produce secondary electrons with a probability
greater than unity, the electron population grows, producing a
so-called ``electron cloud'' (EC\@).  In extreme cases, a large
density of electrons can build up, causing disruption of the beam,
heating of the chamber walls, and degradation of the vacuum.

Electron cloud effects were first observed in the 1960s
\cite{ECLOUD04:9to13}.  A number of adverse effects from EC have been
observed in recent years
\cite{%
PRSTAB6:034402, 
PRSTAB7:094401, 
PRL79:3186, 
PRSTAB9:012801, 
PRL74:5044, 
PRSTAB11:041002, 
PRSTAB5:094401, 
PRSTAB11:010101, 
EPAC08:TUPP043, 
PAC01:TPPH100}. 
Several accelerators were modified to reduce the cloud density
\cite{%
PRSTAB9:012801, 
PAC01:TPPH100, 
PRSTAB11:041002}. 
EC concerns led to EC mitigation features in the design of recent
accelerators \cite{%
EPAC08:TUPP043, 
ECLOUD12:Wed0830} 
and proposed
future accelerators \cite{%
JVSTA30:031602, 
IPAC10:WEPE097, 
EPAC08:MOPP050}. 
 Additional information on EC issues can be found
in review papers such as \cite{ECLOUD04:9to13, ECLOUD12:Wed0830,
  ECLOUD12:Tue1815}.

The Cornell Electron Storage Ring (CESR) provides X-ray beams for
users of the Cornell High Energy Synchrotron Source (CHESS) and serves
as a test bed for future accelerators through the CESR Test
Accelerator program (\CesrTA) \cite{PAC09:FR1RAI02, ECLOUD10:OPR06,
  CLNS:12:2084}.  A major goal of the \CesrTA{} program is to better
understand EC effects and their mitigation.  The EC density is
measured with multiple techniques \cite{PRSTAB17:061001,
  NIMA749:42to46, NIMA754:28to35}.  The effectiveness of several types
of coatings for EC mitigation has been measured on coated and
instrumented chambers \cite{PRSTAB17:061001}.

For a beam emitting synchrotron radiation (SR), three surface
phenomena are important to the build-up of the electron cloud:
photo-emission of electrons; secondary emission of electrons; and
scattering of photons.  Since it is possible for a surface to release
more electrons than are incident, secondary emission can be the
dominant EC growth mechanism.

Surface properties are known to change with time in an accelerator
vacuum chamber: this is referred to as ``conditioning'' or ``beam
scrubbing.''  Beam scrubbing is thought to be due to the removal of
surface contaminants by bombardment from SR photons, scattered
photons, cloud electrons, ions, beam halo, or some combination
thereof.

During the \CesrTA{} program, a system was developed for in-situ
measurements of the secondary electron yield (SEY) as a function of
the energy and angle of the incident primary electrons.  The goals of
the in-situ SEY studies included (i) measuring the SEY of surfaces
that are commonly used for beam chambers; (ii) measuring the effect of
beam conditioning; and (iii) comparing different mitigation coatings.
Samples were made from the same materials as one would find in an
accelerator vacuum chamber, with similar surface preparation
(sometimes called ``technical surfaces'' in the literature).

The effect of exposure to an accelerator environment on the SEY has
been studied by several groups
\cite{JVSTA21:1625to1630, 
EPAC00:THXF102, ECLOUD02:17to28, 
ECLOUD07:72to75, ANTIECLOUDCOAT09:24, 
EPAC02:WEPDO014, PRSTAB14:071001, IPAC11:TUPS028, 
NIMA621:47to56}. 
Systematic errors in SEY measurements and countermeasures have been
studied at SLAC \cite{ECLOUD04:107to111, SLAC:PUB10541}.  In some of
these studies, the samples were installed into the beam pipe for an
extended period and then moved to a laboratory apparatus for SEY
measurements.  At Argonne, the removal of the samples required a brief
exposure to air \cite{JVSTA21:1625to1630}.  At PEP-II, samples were
moved under vacuum using a load-lock system \cite{NIMA621:47to56}.
Studies at CERN and KEK, on the other hand, used in-situ systems, so
that samples did not have to be removed for the SEY measurements
\cite{ECLOUD02:17to28, EPAC02:WEPDO014, ECLOUD07:72to75,
  ANTIECLOUDCOAT09:24}.  The in-situ systems allow for more frequent
measurements with fewer concerns about recontamination before the
measurements, but require a more elaborate system in the accelerator
tunnel.

The SEY apparatus developed for \CesrTA{} was based on the system used
in PEP-II at SLAC \cite{NIMA621:47to56}.  In lieu of the PEP-II
load-lock system, a more advanced vacuum system was designed,
incorporating electron guns for in-situ SEY measurements.  The
measurements at \CesrTA{} are similar to the in-situ measurements at
CERN and KEK, but with several differences: (i) we have studied a
wider variety of materials than measured at CERN; (ii) we have done
more frequent measurements than done at KEK to get a more complete
picture of SEY conditioning as a function of time and beam dose; (iii)
we have measured the dependence of SEY on position and angle of
incidence.  Systems similar to the \CesrTA{} stations were recently
sent to Fermilab for EC studies in the Main Injector
\cite{IPAC12:MOPPC019}.

The \CesrTA{} in-situ samples are typically measured weekly during a
6-hour tunnel access.  The SEY chamber design allows for samples to be
exchanged rapidly; this can be done during the weekly access if
needed.  There are 2 samples at different angles, one in the
horizontal plane, the other 45\degree{} below the horizontal plane, as
was the case at PEP-II.  This allows us to compare conditioning by
direct SR photons in the middle of the horizontal sample versus
bombardment by scattered photons and EC electrons elsewhere.  Because
the accelerator has down periods twice a year, we are able to keep
some samples under vacuum after conditioning to observe the changes in
SEY over several weeks, without exposure to air.

Models have been developed to describe the SEY as a function of
incident energy and angle (e.g.\ \cite{PRSTAB5:124404}).  In the
models, the secondary electrons are generally classified into 3
categories: ``true secondaries,'' which emerge with small kinetic
energies; ``rediffused secondaries,'' with intermediate energies; and
``elastic secondaries,'' which emerge with the same energy as the
incident primary.  The models are used to predict the EC density and
its effect on the beam.  Our in-situ SEY measurement program is
ultimately oriented toward finding more realistic SEY model
parameters, for more accurate predictions of EC effects.

This paper describes the apparatus and techniques developed for the
in-situ SEY measurements.  For clarity, we divide the stages of the
measurement program into three parts, Phase I, Phase IIa, and Phase
IIb.  We describe the in-situ apparatus and basic measurement method
in \autoref{S:AppMethod}.  The Phase I measurement techniques are
summarized in \autoref{S:phasei}.  In Phase II, improvements were made
to the hardware and measurement techniques, as described in
\autoref{S:phaseii}.  The data analysis is discussed in
\autoref{S:analysis}, and examples of results are given in
\autoref{S:exam}.  Additional details on our SEY instrumentation and
methods can be found in a separate report \cite{ARXIV:1407.0772}.
Preliminary SEY results for metals (aluminum, copper, stainless steel)
and EC mitigation films [titanium nitride, amorphous carbon (aC),
  diamond-like carbon (DLC)] can be found in other papers
\cite{CLNS:12:2084, ECLOUD10:PST12, PAC11:TUP230, IPAC13:THPFI087}.

\section{Apparatus and Basic Method\label{S:AppMethod}}

There are two SEY stations to allow exposure of two samples to the
accelerator environment.  The SEY measurements are done in the
accelerator tunnel while the samples remain under vacuum.  To keep the
stations compact enough for deployment in the tunnel, we use an
indirect method to measure the SEY\@.  Our basic measurement method is
the same as was used by SLAC \cite{NIMA551:187to199, NIMA621:47to56,
  ECLOUD04:107to111} and other groups; the instrumentation is the same
as was used at SLAC \cite{NIMA551:187to199}.  We measure the
dependence of the SEY on the (i) incident kinetic energy $K$, (ii)
incident angle $\theta$, and (iii) impact position of the primary
electrons ($\theta$ = angle from the surface normal).  An additional
station outside the tunnel is used for supplementary measurements.

\subsection{Storage Ring Environment}

In CESR, electrons and positrons travel in opposite directions through
a common beam pipe.  Beam scrubbing occurs mostly under CHESS
conditions: a beam energy of 5.3~GeV, with beam currents of $\sim
200$~mA for both electrons and positrons.  The SEY samples are
installed into the wall of a stainless steel beam pipe with a circular
cross-section of inner diameter 89~mm.  The SEY samples are exposed
predominantly to SR from the electron beam, the closest bending magnet
being about 6~m away.  The SEY beam pipe includes a retarding field
analyzer for electron cloud characterization.

Cold cathode ionization gauges are used to monitor the beam pipe
pressure; the closest gauge is about 1~m away.  The base pressure is
generally $\lappeq 1.3 \cdot 10^{-7}$~Pa.  With CHESS beams, the
pressure is typically $\lappeq 6 \cdot 10^{-7}$~Pa after beam
conditioning.

\subsection{In-Situ SEY Stations}

\begin{figure*}[tb]
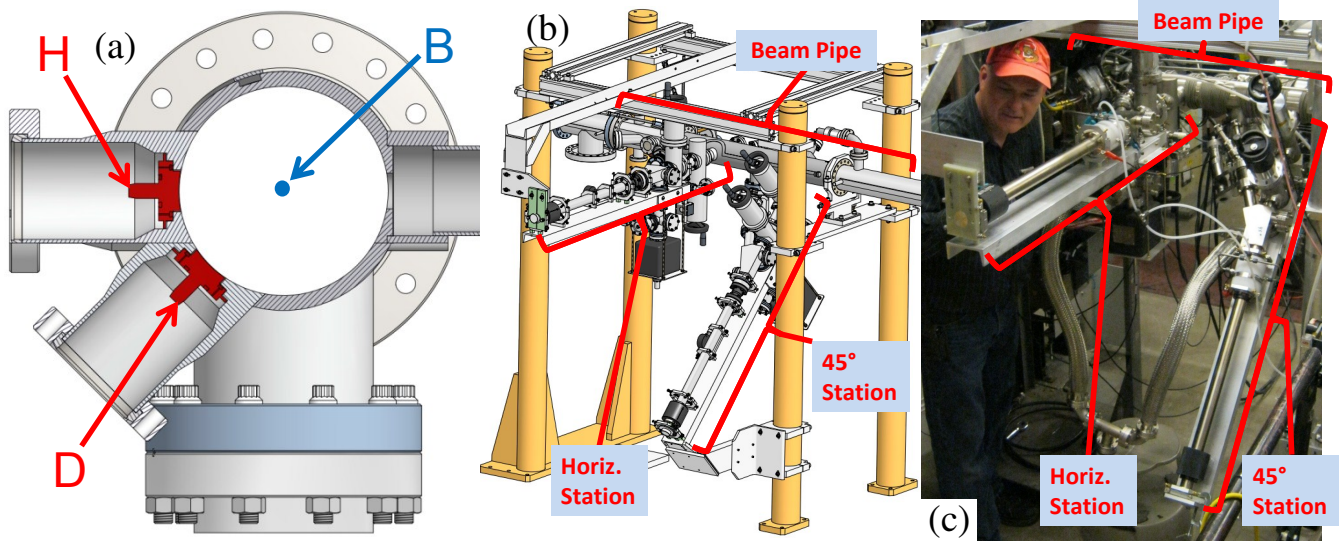

\centering
\makebox[\textwidth]{%
\GRAFheight[\heightone]{250}{280}
\GRAFlabelcoord{50}{250}
\incGRAFlabel{fig01a}{(a)}%
\GRAFheight[\heightone]{440}{540}
\GRAFlabelcoord{45}{500}
\incGRAFlabel{fig01b}{(b)}%
\GRAFheight[\heightone]{420}{540}
\GRAFlabelcoord{0}{4}
\GRAFlabelbox{40}{30}
\incGRAFboxlabel{fig01c}{(c)}%
}\\[-\medspace]

\caption[Beam's eye view, isometric drawing, and photo of SEY
  stations]{(a) ``Beam's eye'' view of the SEY stations (B = beam, H =
  horizontal sample; D = 45\degree{} sample).  (b) Isometric drawing
  and (c) photograph of the SEY stations and beam pipe.  Note that (a)
  does not show the longitudinal separation of $\sim 0.4$~m
  between the samples, though it can be seen in (b) and
  (c).\label{F:sys}}

\end{figure*}

As shown in \cref{F:sys}a, the samples have a curved surface to match
the beam pipe cross-section.  The samples are approximately flush with
the inside beam pipe, with one sample positioned horizontally in the
direct radiation stripe, and the other sample positioned at
$45^\circ$, below the radiation stripe.  Figs.~\ref{F:sys}b and
\ref{F:sys}c show the SEY stations, including the equipment for moving
the samples under vacuum and measuring the SEY.

\begin{figure}[tb]
\centering
\setlength{\unitlength}{\heighttwo/960}%
\begin{picture}(920,960)(0,0)
\put(0,0){\includegraphics[height=\heighttwo]{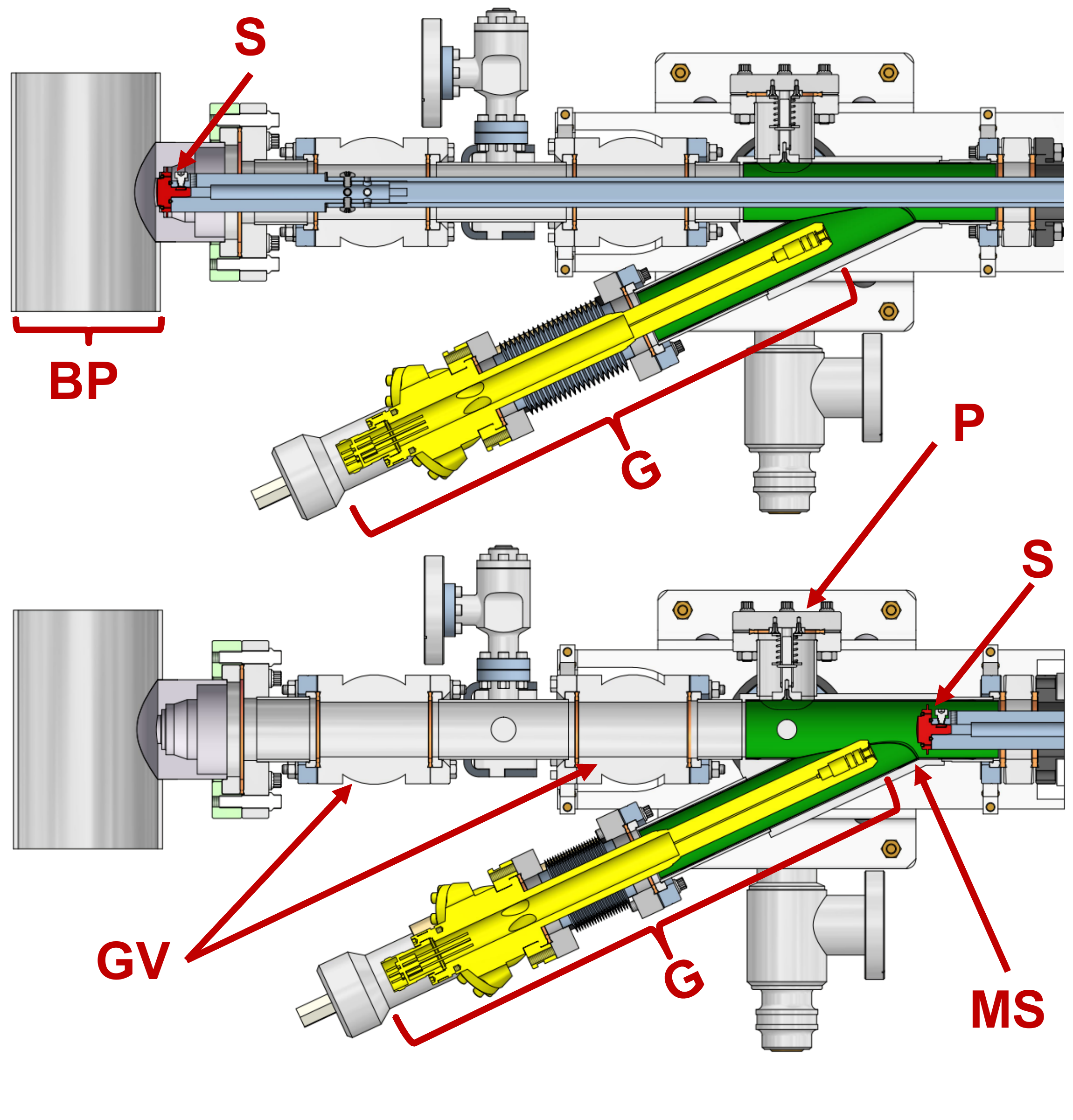}}
\put(800,900){\Large (a)}
\put(725,50){\Large (b)}
\end{picture}\\[-\bigspace]

\caption[Sectional views of SEY station]{SEY station with sample (a)
  inserted in beam pipe and (b) retracted for measurements.  S: sample
  (red); G: electron gun (yellow); MS: magnetic shield (green); BP:
  beam pipe; P: port for sample exchange; GV: gate
  valve.\label{F:arm}}

\end{figure}

More detailed drawings of one SEY station are shown in
\cref{F:arm}.  A custom-designed vacuum ``crotch'' provides an
off-axis port for an electron gun, a pumping port, and a side port for
sample exchange.  The sample is mounted on a linear positioner with a
magnetically-coupled manual actuator.\footnote{Model DBLOM-26,
  Transfer Engineering, Fermont, CA.}  The electron gun is at an angle
of 25\degree{} from the axis of the sample positioner.  The gun is
mounted on a compact linear positioner\footnote{Model LMT-152, MDC
  Vacuum Products, LLC, Hayward, CA.}  so it can move out of the
sample positioner's path when the sample is inserted into the beam
pipe (\cref{F:arm}a).

When the sample is in the beam pipe (\cref{F:arm}a), force is
applied to the actuator to ensure that the sample is well seated.
When the sample is in the SEY measuring position (\cref{F:arm}b),
the gun is moved forward to make the nominal gun-to-sample distance
32.9~mm.  Moving the gun forward allows for a smaller beam spot size
and a larger range of incident angles.

One or both of two gate valves are closed to isolate the CESR vacuum
system from the SEY chambers during SEY measurements.  The pressure
inside the SEY chambers is typically $\lappeq 10^{-6}$~Pa with the
electron guns on (as measured indirectly via ion pump current
read-backs).

\subsection{Samples and Sample Exchange\label{S:sam}}

\begin{figure}[tb]
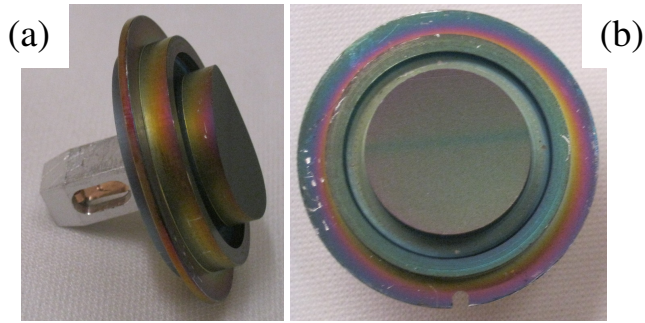

\centering
\makebox[\narrowwidth]{%
\GRAFlabelbox{200}{150}
\GRAFlabelcoord{-100}{825}
\GRAFheight[0.175\textheight]{833}{1000}
\incGRAFboxlabel{fig03a}{(a)}
\GRAFlabelcoord{925}{825}
\GRAFheight[0.175\textheight]{1000}{1000}
\incGRAFboxlabel{fig03b}{(b)}
}\\[-\medspace]

\caption[Photographs of an SEY sample]{Photographs of an SEY sample with a
  TiN coating: (a) side, (b) front.\label{F:grid}}

\end{figure}

The samples, shown in \cref{F:grid}, are machined from bulk material.
They are solvent cleaned without mechanical polishing or etching.
Coatings (if any) are applied after cleaning.

The gate valves allow us to change the samples without venting the beam
pipe.  The SEY chamber has a special port for changing the sample in
the tunnel (see \cref{F:arm}), with a custom-designed patch for the
magnetic shield.  The sample exchange can be done with the flanges
open for only a few minutes.  The ultra-high vacuum recovers
sufficiently to resume measurements 24~hours after venting.  Hence it
is possible to change samples during a scheduled tunnel access over a
CHESS running period.

\subsection{SEY Measurement\label{S:SeyMeas}}

The secondary electron yield is defined as the number of secondary
electrons released from a surface divided by the number of incident
primary electrons.
In terms of current,
\begin{equation}
{\rm SEY} = -\frac{I_s}{I_p} \; ,\label{E:seys}
\end{equation}
where $I_p$ is the primary current and $I_s$ is the secondary current.
The minus sign in \cref{E:seys} is included because the primary and
secondary electrons travel in opposite directions relative to the
sample and hence $I_p$ and $I_s$ have opposite signs.

To measure $I_p$, we fire electrons onto the sample from the gun and
measure the current from the sample with a positive bias voltage.  A
high positive bias, $V_b = +150$~V, is used to recapture
secondaries produced by the primary beam, so that the net current due
to secondaries is zero in the ideal case.

We measure $I_s$ indirectly.  The total current $I_t$ is measured by
again firing electrons at the sample, but with a negative bias
($V_b = -20$~V) to repel the secondaries.  Since $I_t = I_p + I_s$,
\begin{equation}
{\rm SEY} = - \frac{I_t-I_p}{I_p} = 1 - \frac{I_t}{I_p} \; .\label{E:seyt}
\end{equation}
A complication is that some secondaries may hit the wall of the vacuum
chamber and produce additional electrons by secondary emission; hence,
the negative bias should be enough to prevent these electrons from
returning to the sample.  We chose $V_b = -20$~V based on past
measurements at SLAC \cite{ECLOUD04:107to111}.

Some SEY systems include an additional electrode to allow for a more
direct measurement of $I_s$ \cite{IPAC10:TUPD043, ECLOUD04:107to111}.
Our in-situ setup cannot accommodate an extra electrode, so we cannot
use such a method.  We should also note that the positive bias for the
$I_p$ measurement in our indirect method is not able to retain
elastic secondaries, so that the elastic contribution to the SEY is
not fully accounted for, as has been pointed out previously
\cite{ECLOUD04:107to111}.

To measure the SEY, we bombard the sample with electrons, which can
condition it and change the SEY\@.  To observe the effect of SR
photons and EC electrons, it is best to minimize conditioning by the
electron gun \cite{ECLOUD04:107to111}.  A low gun current and a rapid
measurement help with this, but the current must be large enough to
measure and settling times are needed (see \autoref{S:trans}).  As a
result, we ``park'' the beam at a known position with a small beam
spot size when we are not measuring $I_t$.  We will return to the
issue of parasitic conditioning in \autoref{S:charge}.

\subsection{Electron Gun, Gun Deflection, and Spot Size\label{S:gun}}

The electron gun\footnote{Model ELG-2, Kimball Physics, Inc., Wilton,
  NH.}  provides a dc beam of up to 2~keV via thermionic emission,
with electrostatic acceleration, focusing, and deflection.  During the
measurement, we scan the gun energy and the deflection (to measure the
dependence of SEY on incident energy $K$ and incident position,
respectively).  The deflection voltages are scaled with energy to
produce the desired deflection angles.  With the compact in-situ
system, we cannot independently vary the angle and position.  However,
because of the curvature of the sample, scanning the beam spot
vertically changes the position with little change in $\theta$, while
scanning horizontally changes both the position and the angle.

The focusing voltage is adjusted with energy to minimize the beam spot
size for good position and angle resolution.  With the focus adjusted
to minimize it, the estimated beam spot sizes for different energy
ranges are as follows: slightly larger than 1~mm between 20~eV and
200~eV; $\leq 0.75$~mm from 250~eV to 700~eV; about 1.2~mm at 1500~eV
(increasing with energy between 800~eV and 1500~eV)\@.  Separate
collimation measurements were done to find the optimum focus set point
as a function of beam energy and to estimate the spot size
\cite{ARXIV:1407.0772}.

To produce a stable gun cathode temperature and hence a stable $I_p$,
we warm up the cathode before starting the SEY measurement, typically
for 30 to 60 minutes.  During the warm-up period, we set the gun
energy to zero and deflection to maximum to prevent the gun beam from
reaching the sample.  During the SEY scan, $I_p$ changes with energy
and drifts in time (see \cref{F:IpE} below).  The latter is likely due
to imperfect cathode temperature stability after the warm-up (our
choice of warm-up time is constrained by needing to do the
measurements in the available access time).

\subsection{Current Measurements\label{S:elec}}

\begin{figure}[b]
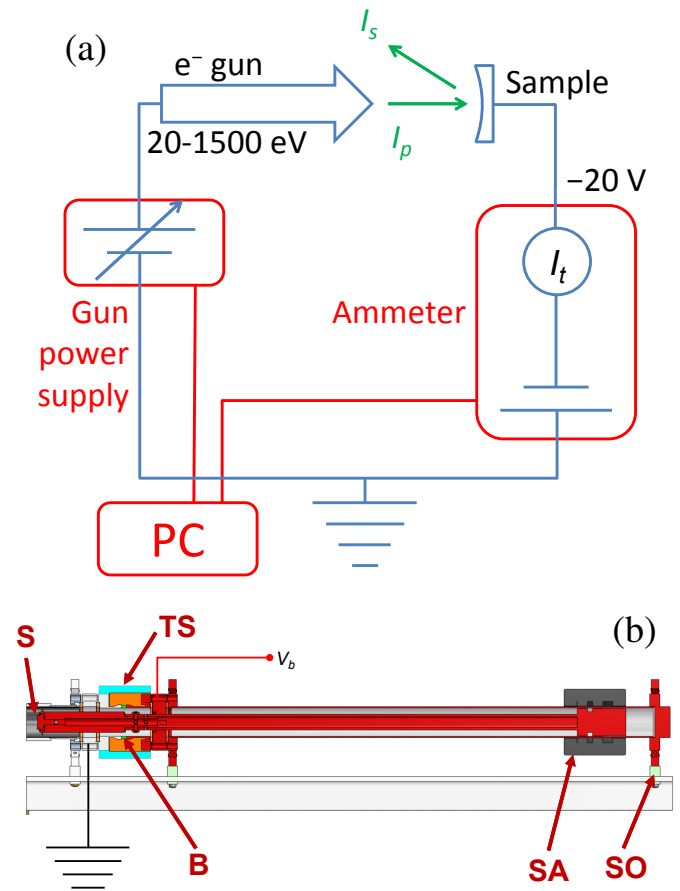

\centering
\GRAFwidth[\narrowwidth]{640}{560}
\GRAFlabelcoord{50}{500}
\incGRAFlabel{fig04a}{(a)}\\[2.5ex]

\GRAFwidth[\widestwidth]{1000}{420}
\GRAFlabelcoord{900}{375}
\incGRAFlabel{fig04b}{(b)}\\[-\medspace]

\caption[Electrical schematic and side view of SEY station]{(a)
  Electrical and data acquisition schematic, showing the sample with a
  negative bias to measure the total current ($I_t$); (b) Side view of
  the SEY station; red indicates portions to which the sample bias is
  applied.  S: sample; B: ceramic break (green); SA: sample actuator;
  SO: stand-off (light green); TS: Teflon shell (light blue) for
  nitrogen gas blanket (orange).\label{F:elec}}

\end{figure}

An electrical schematic of the system is shown in \cref{F:elec}a.
The bias voltage is applied to the sample and positioner arm, which
are separated by a ceramic break\footnote{Model BRK-VAC5KV-275,
  Accu-Glass Products, Inc., Valencia, CA.} from the grounded SEY
chamber (\cref{F:elec}b).  A picoammeter\footnote{Model 6487,
  Keithley Instruments, Inc., Cleveland, OH.} measures the current
from the sample.  Low-noise triaxial cables bring the signals from the
sample positioner arms to the picoammeters.  The picoammeter provides
the biasing voltage: a small shielded circuit connects the bias
voltage from the picoammeter power supply.  The outer conductor of the
triax provides a shield for the signals carried by the middle and
inner conductors.

The sample positioner arms are not electrically shielded.  As a
result, activity that disturbs the air near the SEY stations produces
noise in the current signals.  Hence we minimize personnel activity in
the area when doing SEY measurements.

\subsection{Magnetic Shielding\label{S:shield}}

At low energies ($\lappeq 100$~eV), the electrons can be deflected by
up to a few mm by stray magnetic fields.  A magnetic shield, shown in
green in \cref{F:arm}, is used to mitigate this problem.  The shield
is inside the vacuum chamber and has intersecting tubes for the sample
positioner tube and the electron gun side port.  As described above,
the shield has a patch for the sample exchange port.  The shield was
fabricated from nickel alloy mu-metal sheet of thickness 0.5~mm.
Machining, forming, welding, and final heat treatment were done by a
vendor\footnote{MuShield, Inc., Londonderry, NH.} to our
specifications.  Metal finger stock was spot-welded to the outside of
the shield for electrical grounding.

Measurements with a field probe indicated that the shield reduces the
stray magnetic field to $\lappeq 10$~$\mu$T\@.  To check the
deflection with the shield present, we measured the transmission
through a collimation electrode with a 1~mm slit \cite{ARXIV:1407.0772}.
At each energy, the beam was scanned across the slit using the gun
deflection to determine whether compensation was needed to maximize
the current through the slit.  These measurements confirmed that the
stray magnetic field is well shielded.

\subsection{Data Acquisition}

Each station operates independently with its own electron gun,
picoammeter, and CPU, so that the horizontal and 45\degree{} samples
can be measured in parallel.  The SEY scans are automatic and are
controlled by a data acquisition program (DAQP) implemented in
LabVIEW\@.\footnote{Version 8.2, National Instruments, Austin, TX.}
The DAQP incorporates software from Kimball Physics and Keithley for
control and readout of the electron gun and picoammeter, respectively.
We developed and implemented the algorithms to load gun settings,
pause for the necessary settling times, and record the signals for the
SEY scans \cite{ECLOUD10:PST12}.  Development of the DAQP has been an
important part of our SEY measurement program, resulting in a
relatively sophisticated tool for control of SEY scans.

\section{Measurement Method: Phase I\label{S:phasei}}

The Phase I measurement techniques have been described previously
\cite{CLNS:12:2084, ECLOUD10:PST12, PAC11:TUP230}.  The SEY was
measured on a 3 by 3 grid.  The gun energy was scanned from 20~eV to
1500~eV with a step of 10~eV\@.  The DAQP scanned through the energies
and deflections with a constant sample bias, and repeated the process
after changing the bias.

The first scan was done with $V_b = 150$~V to measure $I_p$,
with gun settings for $I_p \approx 2$~nA\@.  This measurement was done
with the deflection set to park the beam between two grid points to
reduce conditioning at the measurement points.

The second scan stepped through the same gun energies with $V_b =
-20$~V to measure $I_t$.  At each energy, the beam was rastered across
all 9 grid points while the DAQP recorded the current for each.  The
$I_t$ scan took $\sim 15$ minutes.

\begin{figure}[htb]
\centering
\includegraphics[width=\widewidth]{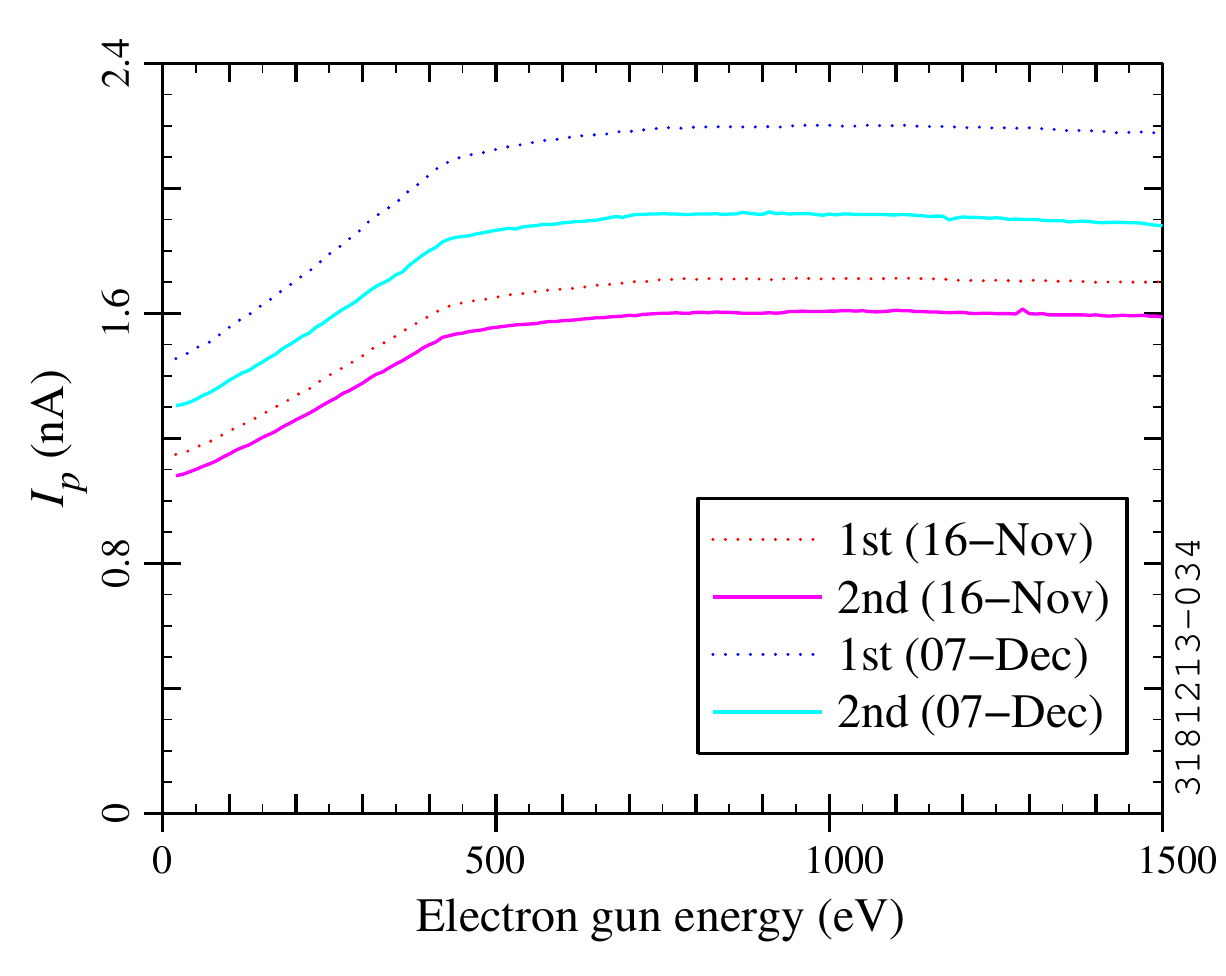}\\[-\bigspace]

\caption[Repeated scans of $I_p$ as a function of energy]{Repeated
  scans of $I_p$ as a function of gun energy (horizontal aC sample,
  2010).  The measurements were done before (dotted curve) and after
  (solid curve) an $I_t$ scan.  Ideally, $I_p$ should be constant at
  2~nA.\label{F:IpE}}

\end{figure}

As indicated above, the gun current varies with gun energy and drifts
in time.  To minimize the error due to current drift, we did a second
$I_p$ scan after the $I_t$ scan.  The first and second $I_p$ values
for a given energy were averaged for calculation of the SEY as a
function of energy and grid point.  \cref{F:IpE} shows examples of
``before $I_t$'' and ``after $I_t$'' scans of $I_p$.

\section{Measurement Method: Phase II Improvements\label{S:phaseii}}

\begin{figure}[htb]
\centering
\includegraphics[width=\widewidth]{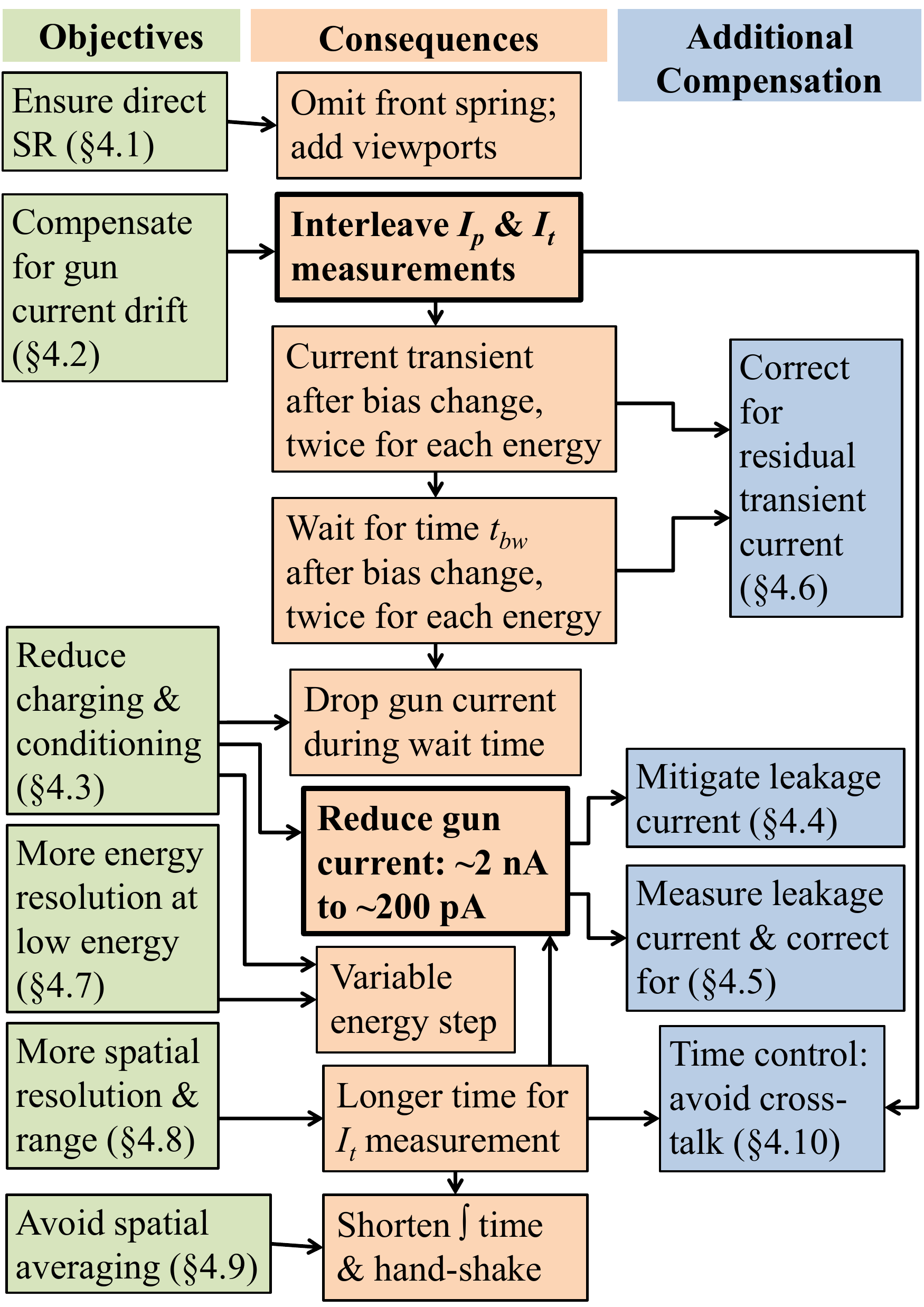}\\[-\medspace]

\caption[Flowchart of modifications to SEY measurement
  method]{Flowchart showing the modifications to the SEY measurement
  method for Phase II and the interrelationships amongst various
  modifications.  Changes with major ramifications are
  highlighted in bold type.\label{F:flowchart}}

\end{figure}

Our experience in Phase I led to iterations in the measurement method.
Modifications for Phase II are described in this section, as outlined
in \cref{F:flowchart}; a time-line can be found in \ref{S:hist}.  The
measurement hardware and methods have been relatively stable since the
start of Phase IIb in August 2012.

Our final procedure for SEY measurements in Phase IIb is as follows:
(i) do a leakage scan (the purpose of which will be described in
\autoref{S:LCmeas}); (ii) warm up the electron guns for 30 to 60
minutes and then set $I_p \approx 200$~pA (\autoref{S:gun}); (iii) do
an SEY scan; (iv) repeat the leakage scan.  The leakage scan takes 40
minutes and the SEY scan takes 110 minutes.  Including sample
transfer, the full measurement takes about 5 hours.  This requires us
to measure the samples in parallel rather than sequentially, since the
access time is typically 6 hours.

\begin{figure}[htb]
\centering
\includegraphics[width=\widewidth]{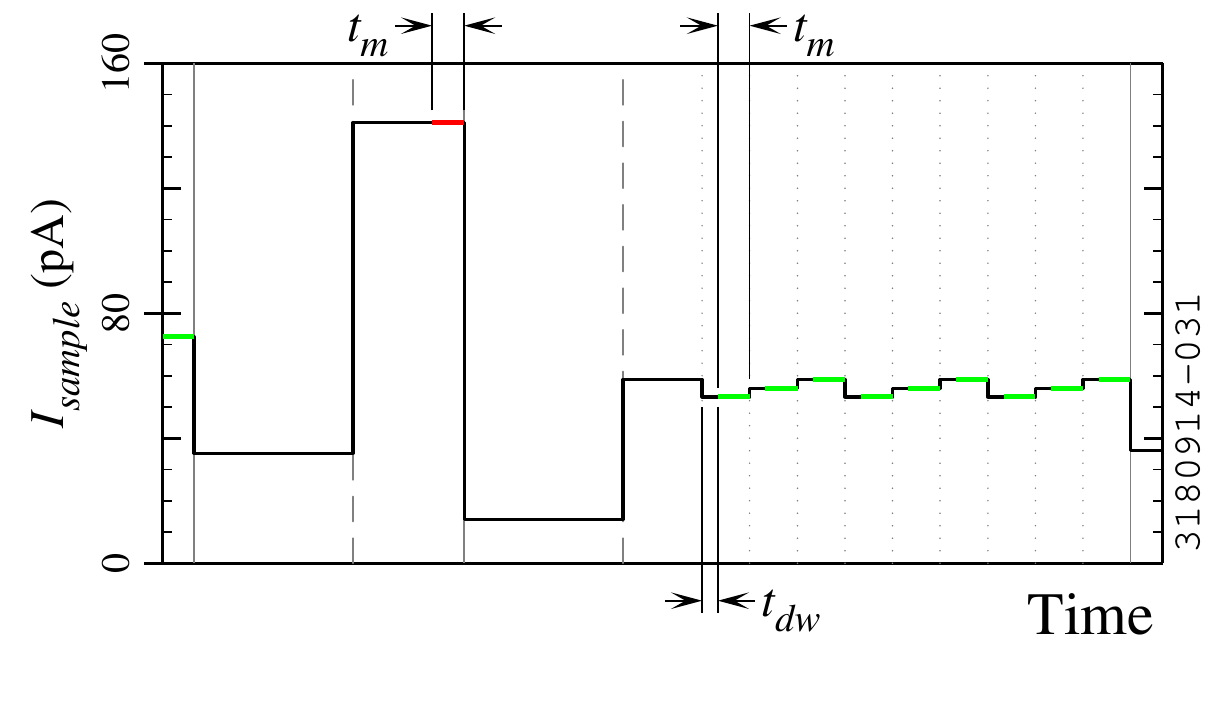}\\[-\biggerspace]

\caption[Zoomed-in timing schematic for sample current]{Zoomed-in
  timing schematic of sample current for SEY scans in Phase IIb, with
  the time intervals for averaging $I_p$ (red) and $I_t$
  (green).\label{F:SEYtimezoom}}

\end{figure}

The final timing algorithm for SEY scans is significantly different
from Phase I, as shown in \cref{F:SEYtime}.  The focus and
deflection are adjusted with energy (\cref{F:SEYtime}b-d); positive
and negative bias are applied at each energy (\cref{F:SEYtime}f)
for current measurements (\cref{F:SEYtime}g; a zoomed-in version is
shown in \cref{F:SEYtimezoom}) with the deflection rastered over
multiple grid points (\cref{F:SEYtime}c-d).  (For clarity, the time
axis is not to scale and a simple 3 by 3 grid is shown.)  Features of
\cref{F:SEYtime} will be discussed further in this section.
\autoref{T:SEYtimepar} gives the final timing parameters.

\begin{table}[htb]
\centering
\caption{Timing parameters for Phase IIb SEY
  scans.\label{T:SEYtimepar}}
\vspace*{\smallspace}
\begin{tabular}{lcl} \hline
Symbol & Value & Description \\ \hline
$t_m$ & $\sim 250$ ms & average and read out current\\
$t_{dw}$ & 50 ms & wait after setting gun deflection\\
$t_{cw}$ & 10 s & wait after setting gun current\\
$t_{bw}$ & 60 s & wait after setting bias\\ \hline
\end{tabular}

\end{table}

\begin{figure}
\centering
\GRAFwidth[0.91\narrowwidth]{490}{205}
\GRAFlabelcoord{475}{155}
\begin{tabular}{c}
\incGRAFlabel{fig07a}{(a)}\\[-3.5ex]
\incGRAFlabel{fig07b}{(b)}\\[-3.5ex]
\incGRAFlabel{fig07c}{(c)}\\[-3ex]
\incGRAFlabel{fig07d}{(d)}\\[-3ex]
\incGRAFlabel{fig07e}{(e)}\\[-3ex]
\incGRAFlabel{fig07f}{(f)}\\[-3ex]
\incGRAFlabel{fig07g}{(g)}\\[-1.5ex]
\end{tabular}\\[0ex]

\caption[Timing schematic for SEY scans]{Timing schematic for SEY
  scans in Phase IIb: (a) gun energy, (b) focus, (c) horizontal and
  (d) vertical deflection, (e) gun emission current, (f) sample bias,
  and (g) sample current as a function of time for 2 iterations
  (75~eV, 100~eV) in the energy scan.  In (g), the averaging of $I_p$
  is in red and the averaging of $I_t$ is in green ($n_x$ and $n_y$ =
  number of horizontal and vertical grid points; $n = n_x n_y$).  Gray
  lines: bias change (solid); increase in the gun current from the
  standby value to the full value (dashed); or deflection change
  (dotted).
\label{F:SEYtime}}

\end{figure}

\subsection{Ensuring Direct Photon Bombardment\label{S:protrude}}

The SR photons are nearly tangent to the beam pipe wall, so a sample
recessed by $\geq 0.1$~mm does not receive any direct photons.  As a
result, in Phase I, we were unsure whether the samples were bombarded
by direct photons; little difference was observed between the
horizontal and 45\degree{} samples.  For Phase II, we took steps to
ensure direct SR bombardment \cite{ARXIV:1407.0772}, and significant
differences were observed in the early conditioning.

\subsection{Mitigation of Electron Gun Current Drift\label{S:Istab}}

As discussed in \autoref{S:phasei}, $I_p$ changes slowly with time.
For the Phase I measurements on Al and aC-coated samples, the ``before
$I_t$'' and ``after $I_t$'' measurements of $I_p$ differed by about
8\% on average and by about 16\% in the worst case.  The first pair of
measurements in \cref{F:IpE} show typical reproducibility.

To reduce the systematic error due to $I_p$ drift, a new measurement
procedure was developed for Phase II in which $I_p$ and $I_t$
measurements are interleaved.  As shown in \cref{F:SEYtime}, the Phase
II procedure is to set the gun energy, apply a positive bias to the
sample, move the beam to the parking point and wait for the current to
stabilize, measure $I_p$ at one grid point, apply a negative bias,
park the beam and wait for the current to stabilize, measure $I_t$ for
all desired grid points, and then proceed to the next energy.  We wait
for a time $t_{bw} = 60$~s after a bias change
(\autoref{T:SEYtimepar}), a compromise between the need for a short
measurement time and the need to allow the transient current to
diminish (\autoref{S:trans}).  The longer waiting time required us to
reduce the number of energy steps (\autoref{S:EngRes}).

With the Phase II method, we estimate that the error in the current
measurements due to gun current drift is $\lappeq 2$\%.

\subsection{Reduction of Charging and Parasitic Conditioning\label{S:charge}}

\begin{figure}[htb]
\centering
\includegraphics[width=\widewidth]{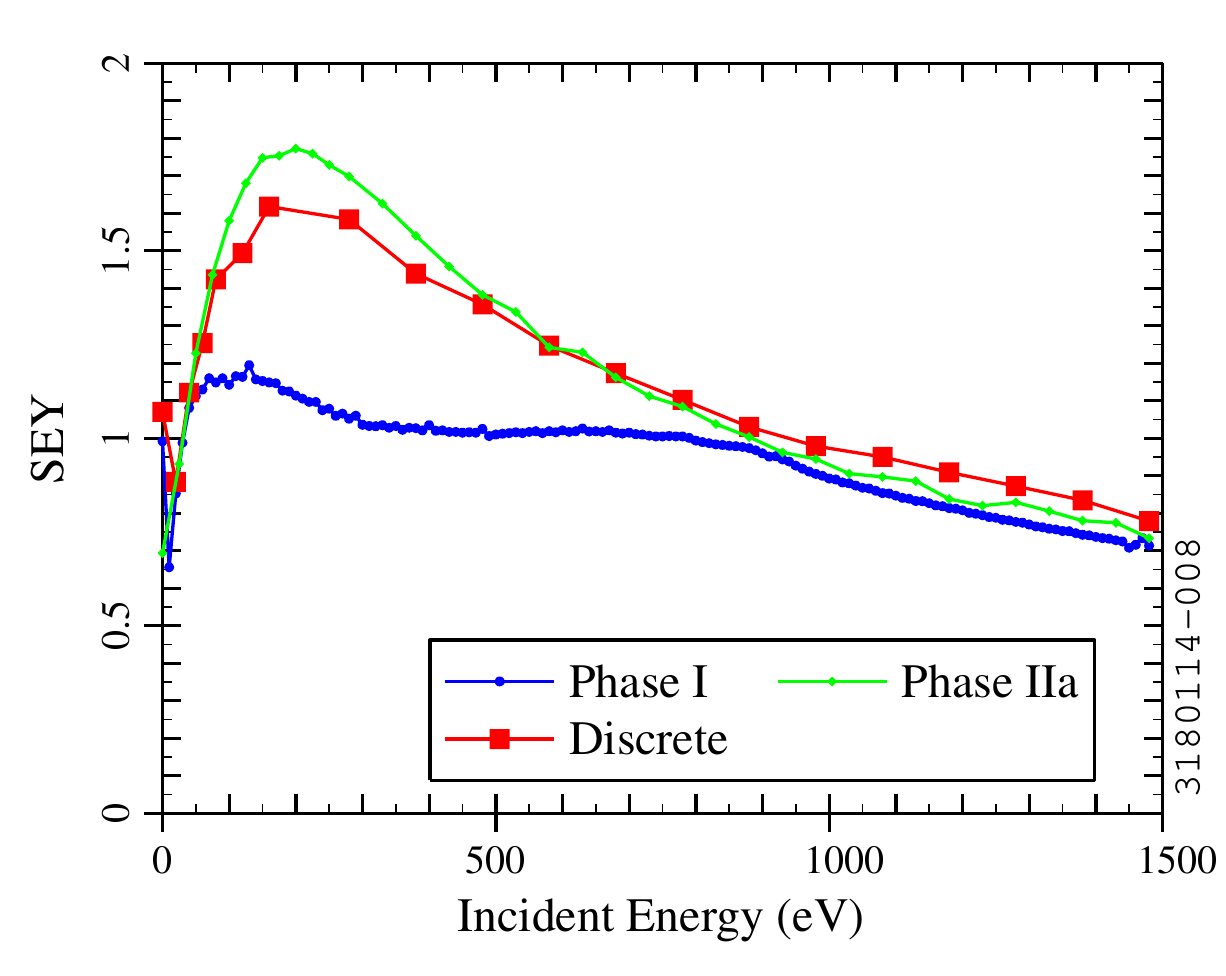}\\[-\bigspace]

\caption[SEY as a function of energy for a DLC sample using different
  measurement methods]{SEY as a function of incident electron energy
  for a DLC-coated Al sample (middle grid point, $\theta =
  25\degree$).  Blue: Phase I method ($I_p \sim 2$~nA, 5 seconds for
  each energy, 9 grid points).  Red: ``discrete'' scan (large energy
  step, $I_p \sim 0.5$~nA, $\sim 3$ minutes waiting period with the
  beam parked away from the measurement point; 1 grid point).  Green:
  Phase IIa method ($I_p < 0.2$~nA, 9 grid points, beam parked away
  from measurement point).\label{F:dlc}}

\end{figure}

Initial SEY measurements on samples with DLC coatings were done in
2011 in the off-line station.  A Phase I measurement on DLC is shown
in \cref{F:dlc} (blue circles).  The SEY curve appears distorted.  We
suspected that the distortion was due to charging of the surface by
the electron beam, presumably owing to insulator-like properties of
the DLC layer.  Similar effects have been reported for other materials
such as MgO \cite{APPSURFSCI111:259to264}.

To test the charging hypothesis, we remeasured the SEY with a long
wait time ($\sim 3$ minutes) between energy steps to allow the surface
to discharge, and with a smaller current ($I_p \sim 0.5$~nA instead of
$\sim 2$~nA) to reduce the supply of charge to the sample.  The beam
was parked away from the measurement point while waiting.  This
resulted in a significant increase in SEY, as shown in \cref{F:dlc}
(red squares).  The new curve is closer to what is measured for other
materials, and is more consistent with other DLC measurements
\cite{SHINKU48:118to120, ANTIECLOUDCOAT09:24, IPAC11:TUPS028}.

The DLC results motivated us to reduce the gun current: in Phase II,
we used $I_p \sim 0.2$~nA\@; a side benefit was to reduce parasitic
conditioning, which, as discussed in \autoref{S:SeyMeas}, should be
minimized.  A complication is that, in Phase II, we switched the bias
to measure $I_p$ and $I_t$ at each energy (\autoref{S:Istab}), with an
added settling time (\autoref{S:trans}).  The longer wait
increased the integrated current per energy step; to shorten the
measurement time and reduce charging and conditioning, we adjusted the
energy segmentation (\autoref{S:EngRes}).  The net result was an
increase in the integrated flux for the parking point and a decrease
in integrated flux for other grid points.  A measurement on the same
DLC sample using the Phase IIa method is included in \cref{F:dlc}
(green diamonds).  The differences between the discrete scan and the
Phase IIa scan are mainly due to the leakage correction
(Sections~\ref{S:LCmeas} and \ref{S:LCC}) included in the Phase IIa
case.

In Phase IIb, an additional improvement was introduced: we decreased
$I_p$ by a factor of $\sim 4$ while waiting for the sample current to
stabilize after a bias change.  We return to the nominal gun
parameters for a time $t_{cw} = 10$~s to allow the gun current to
stabilize before the measurement (\cref{F:SEYtime}).  The gun current
modulation reduces the dose to the parking point by a factor of $\sim
3.5$.  Though different in the details, our modulation method is
conceptually similar to previously-used techniques for insulating
materials (see \cite{APPSURFSCI111:259to264}, for example).

With Phase IIb parameters, one SEY scan produces an integrated
electron flux of $\sim 0.8$~$\mu$C/mm$^2$ for the parking point and
$\lappeq 12$~nC/mm$^2$ for the other grid points.  In past studies on
electron gun conditioning by other groups, the peak SEY decreased by
$\lappeq 10$\% for doses of order 1~$\mu$C/mm$^2$ for Cu
\cite{EPAC00:THXF102, EPAC02:WEPDO014}, TiN \cite{NIMA551:187to199},
and Al \cite{JVSTA23:1610to1618}.  Based on this, we would expect to
see some conditioning at the parking point.  However, there may be
less conditioning in our Phase II SEY scans because the electron
energy is low for much of the scan, and it has been found that
conditioning is less efficient at low energies \cite{PRL109:064801}.

We did not see a significant difference in the parking point's SEY for
Cu, stainless steel, TiN, or Al in Phase II\@.  As noted above, DLC is
more susceptible to charging.  Off-line measurements on an
unconditioned DLC sample with Phase IIb parameters showed a decrease
in the measured SEY at the parking point with current modulation
($\sim 7$\%) and a larger decrease without current modulation ($\sim
24$\%).  On the other hand, a conditioned and air-exposed DLC sample
did not show a difference in measured SEY at the parking point.
Additional measurements are being done in the off-line station to
better quantify the susceptibility to charging and conditioning and
check the reproducibility of our observations.

\subsection{Leakage Current: Mitigation\label{S:LCmit}}

Ideally, the picoammeter measures only the current due to primary and
secondary electrons.  In reality, because the insulators are
imperfect, additional current (``leakage current'') flows through the
picoammeter to ground when the sample is biased.  The leakage current
should be a small fraction of $I_p$ to avoid systematic errors in the
calculated SEY \cite{ECLOUD04:107to111}.  In Phase I ($I_p \sim
2$~nA), no leakage corrections were applied.  Because the relative
contribution from the leakage current increases as the gun current
decreases, the leakage current was investigated while preparing for
Phase II measurements with $I_p \sim 0.2$~nA.

Measurements indicated that the leakage current was strongly
correlated with the ambient humidity.  At high humidity, we found that
the leakage could be as high as several nA (hence exceeding $I_p$) and
could vary significantly in the time needed for an SEY scan, which
could produce large systematic errors.

We took steps to minimize the leakage paths in the measurement circuit
\cite{ARXIV:1407.0772}.  After these modifications, the main leakage paths
were found to be the insulating stand-offs and the ceramic break
(shown in green in \cref{F:elec}b).

The decrease in resistivity of insulators due to moisture has been
documented in the literature in the past century (see, e.g.,
\cite{BULLBSUS11:359to420, JAP17:318to325, IEEEEI11:76to80}).  In a
humid environment, current is conducted along insulator surfaces,
where there is a layer of moisture from the ambient air.  These
considerations led us to a redesign: (i) the original G10 stand-offs
were replaced by similar parts with a smoother surface finish, more
careful cleaning, and blind holes in lieu of through holes; (ii) a
nitrogen gas ``blanket'' was made to isolate the ceramic break from
the air.  A Teflon tube (shown in blue in \cref{F:elec}b) was attached
to the grounded side of the ceramic break, with a small gap on the
biased side to avoid adding another leakage path (the blanket region
is shown in orange in \cref{F:elec}b).  We used a steady flow of N$_2$
gas (about 2.5 SCFH $\approx$ 20~mL/s per station) to establish the
blanket.  The gas source is boil-off from the building's liquid N$_2$
storage Dewar.

At high humidity, the nitrogen blanket alone did not produce a low and
stable leakage current; we had to first warm the ceramic with a heat
gun to remove the existing moisture.  At low humidity, the leakage
currents with and without gas flow were comparable.  See \ref{S:humid}
for more information.

After the modifications to the system, the leakage current was
$\lappeq 30$~pA at $V_b = 150$~V\@.  This corresponds to an error of
$\lappeq 14$\% in the $I_p$ measurement for Phase II parameters (not
including the transient contribution, which is discussed in
\autoref{S:trans}).  Repeated measurements indicated that the leakage
current still varies over time, even with the gas blanket.  The
variation can be as much as a factor of 2 over long periods; see 
\ref{S:LCtrend} for more information.

\subsection{Leakage Current: Measurement\label{S:LCmeas}}

Since the leakage current is not negligible relative to $I_p$ and
varies over time even with mitigation, we measure the leakage current
prior to each SEY scan.  The leakage scan is done with the same
procedure as the SEY scan, but with the gun turned off.  We repeat
several iterations of positive and negative sample bias to allow the
current to stabilize; however, we perform fewer iterations for the
leakage scan (16) than for the SEY scan (44).  The measured values of
$I_p$ and $I_t$ are corrected by subtracting the measured leakage
current for the corresponding bias before calculating SEY
(\autoref{S:LCC}).

Time permitting, a second leakage scan is done after the SEY scan to
quantify the leakage current stability.  Typically, the leakage
currents before and after the SEY scan agree within $\pm 2$~pA for
$V_b = +150$~V and within $\pm 0.5$~pA for $V_b = -20$~V\@.  Hence we
estimate that the leakage current drift contributes an error in the
corrected currents of $\sim 1$\% of $I_p$.

\subsection{Transient Current: Mitigation\label{S:trans}}

A bias change produces a transient in the sample current due to the
stray capacitance of the system and the response of the picoammeter.
The stray capacitance includes a contribution from the triaxial cable
and the SEY station, whose biased positioner arm is in proximity to
the grounded surroundings (\cref{F:elec}b).

The transient current peaks at about 0.5~nA (hence exceeding $I_p$ for
Phase II) with a decay time of order 30~s (examples are included in
\ref{S:derive}).  Ideally, one would wait for the current to reach its
equilibrium value before starting the measurement.  This could be done
in Phase I, since the bias voltage was switched infrequently (twice
per SEY scan).

On the other hand, with the Phase II procedure to mitigate the gun
current drift, the bias is switched twice for each energy
(\cref{F:SEYtime}f), making a long wait time after each bias change
impractical.  Hence a compromise solution was necessary: waiting for
time $t_{bw} = 60$~s after a bias change, reducing the number of
energy steps (\autoref{S:EngRes}), and correcting for the residual
effects from the transients.  Because the leakage scans described
above are done while switching the bias with the same timing algorithm
as is used for the SEY scans, the correction for the leakage current
also corrects for the residual transient current, which is about 4\%
of $I_p$ for the Phase II parameters.

The transient response produces a change in the leakage current over
the time required to measure all of the grid points in the double scan
(see \cref{F:LCT} below).  We record the time stamp along with the
current, as the time elapsed since a bias change varies due to
skipping of grid points (\autoref{S:pts}) and compensation of the
waiting time (\autoref{S:cross}).  A time dependence is included in
the leakage correction to account for the change in current during the
$I_t$ measurements (\autoref{S:LCC}).

\subsection{Energy Resolution and Segmentation\label{S:EngRes}}

Phase I measurements were done with a fixed energy step of 10~eV\@.
Because low-energy electrons are important to the build-up of the
electron cloud, a smaller step was of interest for low energies.  In
Phase II, both $I_p$ and $I_t$ were measured in a single energy scan
(\autoref{S:Istab}), with longer waiting times at each energy
(\autoref{S:trans}).  To keep the overall measurement time within the
schedule constraints, the energy step was increased for high energies
(further motivated by the need to minimize charging and parasitic
conditioning, as discussed in \autoref{S:charge}).  The final Phase II
method was to segment the energy range into 5 pieces, with an initial
step of 1~eV, a final step of 75~eV, and a total of 44 energies
\cite{ARXIV:1407.0772}.

\subsection{Improved Spatial Resolution and Range\label{S:pts}}

\begin{figure}[b]
\centering
\includegraphics[width=\widewidth]{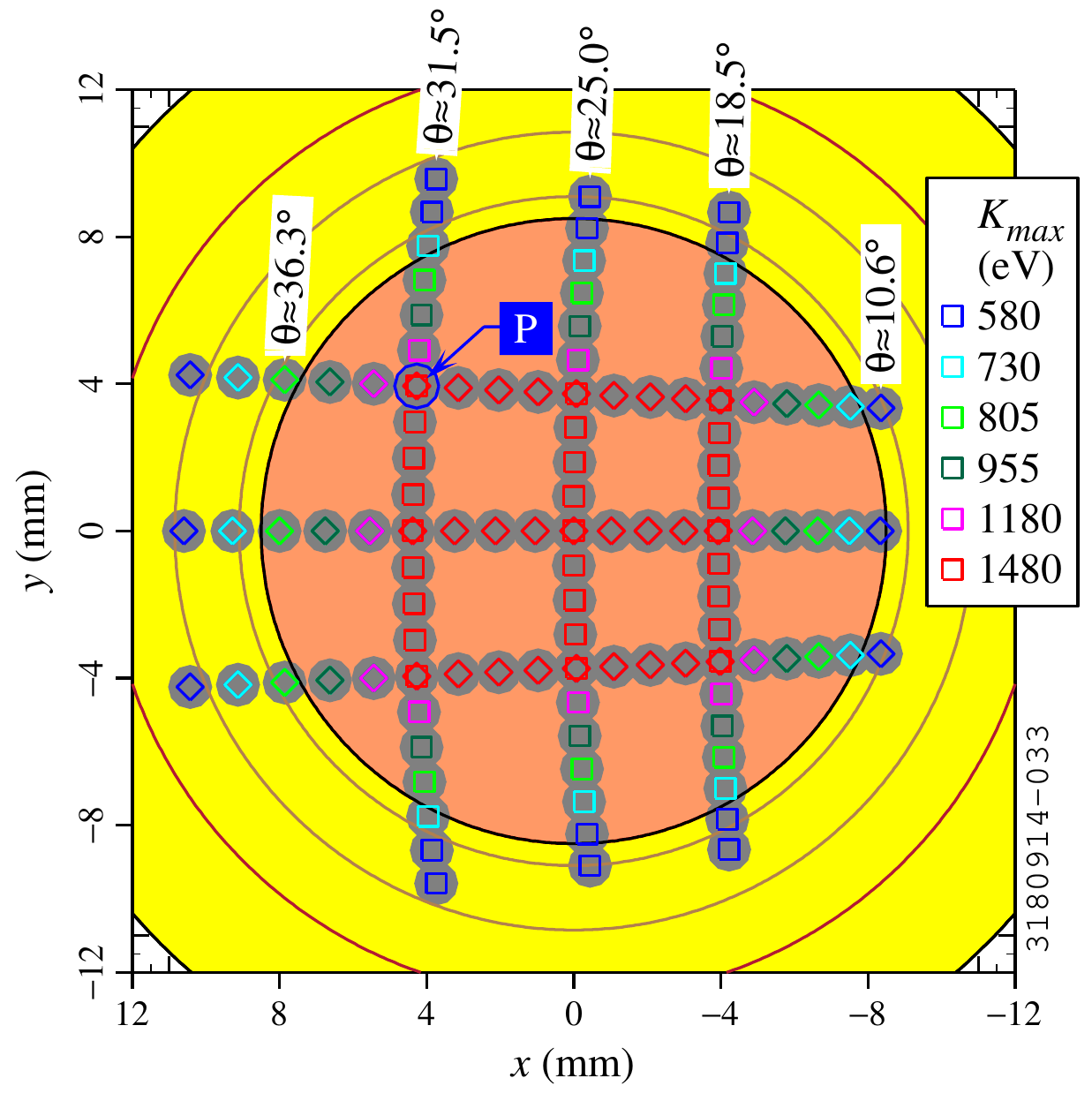}\\[-\bigspace]

\caption[Grid points for double scans]{Grid points for double scans
  ($x$ and $y$ are the horizontal and vertical distance from the
  middle of the sample, respectively).  Legend: maximum incident
  energy measured for each grid point.  Squares: first array;
  diamonds: second array.  Solid gray circles: estimated beam spot
  size at high gun energy.  The approximate incident angle ($\theta$)
  is indicated for selected grid points.  Orange: sample face; yellow:
  sample shoulder.  P: parking point.\label{F:GridSamXY}}

\end{figure}

As discussed above, the Phase I measurements were done over a 3 by 3
grid.  We implemented scans with increased range and resolution in
Phase II to get a better picture of the SEY's dependence on position
and angle.  A uniformly-spaced grid with high resolution and full
range is not practical for weekly measurements, so we scan over 3
horizontal segments and 3 vertical segments only.  (Occasionally,
scans with high resolution and full range are done when additional
time is available.)  The grid points are shown in
\cref{F:GridSamXY}.  One complication is that the largest
deflections cannot be reached at high energies, because the gun's
deflecting electrodes are limited to $\pm 150$~V\@.  The colors in
\cref{F:GridSamXY} indicate the maximum energy measured for
each grid point.  The DAQP skips out-of-range points, which
complicates the timing, as will be discussed in \autoref{S:cross}.

For simplicity, the grid point layout shown in \cref{F:GridSamXY}
is measured using two arrays of gun deflections.  As a result, 9 of
the grid points are measured twice (the repeated points coincide with
the 3 by 3 grid of Phase I).  This provides additional information
about systematic and statistical errors.

For simplicity, the gun deflection is varied linearly between grid
points, leading to a constant increment in the tangent of the
deflection angle.  Because the gun axis is 25\degree{} from the sample
axis, the grid point spacing is not exactly left-right symmetric.
Moreover, the sample face is curved, which shifts the grid points
slightly near the upper and lower edges of the sample.  These
asymmetries in the grid layout can be seen in \cref{F:GridSamXY}.

In \cref{F:GridSamXY}, the gray circles indicate the estimated beam
size at 1500 eV (not accounting for possible distortion in the beam
spot for large deflecting angles).  There is overlap between adjacent
points over most of the sample.  The estimated beam spot size is
smaller at intermediate energies (\autoref{S:gun}); for the smaller
spot size, none of the grid points overlap.

\subsection{Spatial Resolution: Time Control and Hand-Shaking\label{S:hshake}}

In Phase I, we unintentionally used incompatible timing parameters for
the picoammeter and the DAQP, such that $I_t$ measurements were
averaged over multiple grid points.  In Phase II, we decreased the
averaging time and implemented a ``hand-shaking'' algorithm: after
setting the gun parameters, the DAQP waits for the settling time
$t_{dw}$, and then instructs the picoammeter to clear its buffer,
average the current, and return the averaged value.  The DAQP waits
for the picoammeter's value before proceeding, which avoids
unintentional averaging.  With the final Phase II method, the current
is averaged over $\frac{1}{6}$~sec, and the net measurement time per
grid point is about 0.3~sec.

\subsection{Time Control: Cross-Talk Avoidance\label{S:cross}}

As discussed in \autoref{S:elec}, shielded cables connect the sample
positioner arms to the picoammeters, but the positioner arms are not
electrically shielded.  The stations are relatively close together
($\sim 0.4$~m apart at the beam pipe).  As shown in
\cref{F:SEYtime}f, the Phase II SEY scan algorithm requires two
steps in the bias voltage for each energy.  We observed that a bias
change on one sample produces a spike in the measured current of the
other sample.  With Phase II parameters, the current perturbation can
be up to $\sim 50$\% of $I_p$.  If the bias of one sample is changed
while the current of the other sample is being measured, this can
produce a noise spike in the measured SEY.

To avoid noise spikes, we implemented a delay between the scan start
times.  In Phase IIb, the bias wait time is 60~s and the $I_t$
measurements take $\sim 35$~s, allowing for a timing margin of
$\sim 12$~s with a start delay of 47~s between systems.

Even with start time control, spikes still occurred occasionally.
Further investigation indicated that the time to measure one grid
point is sometimes much longer ($\sim 1$~s) than nominal (0.3~s).  When
the longer delays are random, there is little cumulative effect.  With
the large number of Phase II grid points, a cluster of longer delays
sometimes occurs for one system, accumulating enough time difference
to produce cross-talk again.

To eliminate the cross-talk problem reliably, we modified the timing
algorithm.  The DAQP uses the wait times and expected measurement time
per point to predict the overall time per energy step.  After each
energy step, the DAQP checks the time elapsed.  In the next energy
step, it adjusts the wait time to compensate for the actual time of
the previous step being different from the desired time.  This
prevents timing variations from accumulating a large time offset.  It
has the side effect that the wait time varies from one energy to
another; this is taken into account in the data analysis, as discussed
in \autoref{S:LCC}.

A complication is that the number of grid points decreases for $K >
580$~eV (\autoref{S:pts}).  Hence the DAQP must account for skipped
points when calculating the expected time.  As long as their start
times have the appropriate offset, the 2 systems remain synchronized
as the time per iteration decreases.

\section{Data Analysis\label{S:analysis}}

The SEY is calculated from $I_p$ and $I_t$ using \cref{E:seyt}.  The
gun energy is corrected to account for the electrostatic deflection
and the sample bias.  The measured currents are corrected to account
for the leakage and transient current.

\subsection{Momentum Corrections\label{S:BiasCor}}

Paired electrodes at the exit of the gun deflect the electrons by the
desired horizontal and vertical angles ($\alpha_x$, $\alpha_y$).
Because the kicks are electrostatic, they also change the electrons'
kinetic energy.  In the non-relativistic case, the kinetic energy
$K_g$ of the electrons is related to the set point value $K_{gsp}$ via
\begin{equation}
K_g = K_{gsp} [1 + \tan^2(\alpha_x) + \tan^2(\alpha_y)] \; .
\end{equation}
For Phase IIb parameters, the energy correction is $\leq 9.5$\%.

The SEY is a function of the kinetic energy $K$ and angle $\theta$ of
the incident primary electron.  Because the sample is biased, $K$ is
shifted from the gun energy ($K_g$) by the bias voltage ($V_b$):
\begin{equation}
K = K_g + q_e V_b\label{E:K} \; ,
\end{equation}
where $q_e$ is the electron charge magnitude.  Hence the incident
energy is smaller than $K_g$ by 20~eV when we measure $I_t$ and larger
by 150~eV when we measure $I_p$.  Ideally, the negative bias repels
all of the secondary electrons from the sample, while the positive
bias prevents the escape of any secondaries.  In the ideal case,
assuming that the intrinsic $I_p$ is independent of $V_b$, we may use
$V_b = -20$~V in \cref{E:K} to calculate the appropriate incident
energy $K$ associated with the measured SEY.

Because the primary electrons' incident angle is not normal to the
sample ($\theta \neq 0$), the sample bias can also shift the incident
angle and impact position.  At present, our data analysis does not
account for these effects.  As a result, the reader should be cautious
about making inferences about the SEY for $K \lappeq 100$~eV based on
our measurements \cite{ARXIV:1407.0772}.

\subsection{Correction for Leakage and Transient Current\label{S:LCC}}

In Phase II, we mitigated (\autoref{S:LCmit}) and measured
(\autoref{S:LCmeas}) the leakage current, including the transient
contribution (\autoref{S:trans}).  To properly correct the SEY, we
developed a model for the leakage current that includes the transient
contribution, as described in \ref{S:derive}.

\begin{figure}
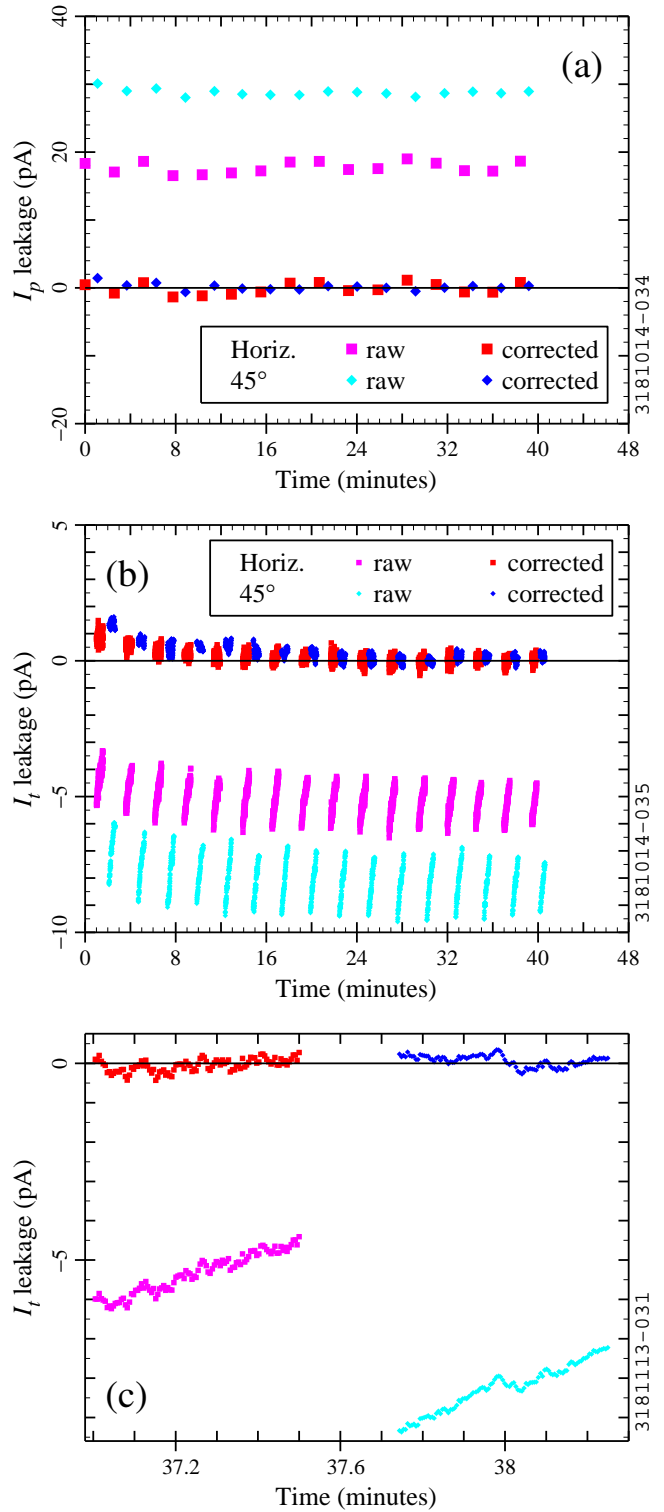

\centering
\GRAFwidth[\narrowwidth]{490}{390}
\GRAFoffset{-65}{-65}
\GRAFlabelcoord{350}{255}
\incGRAFlabel{fig11a}{(a)}\\[-2ex]
\GRAFlabelcoord{15}{255}
\incGRAFlabel{fig11b}{(b)}\\[-2ex]
\GRAFlabelcoord{15}{25}
\incGRAFlabel{fig11c}{(c)}\\[-\medspace]

\caption[Example of measurements of leakage current as a function of
  time while switching bias]{Example of measurements of leakage
  current as a function of time while switching the bias voltage
  (stainless steel samples, Aug 2012).  (a) Leakage measurements with
  $V_b = +150$~V for $I_p$ correction (1 point per iteration); (b)
  leakage measurements with $V_b= -20$~V for $I_t$ correction (120
  points per iteration); (c) same as (b), with zoomed-in view of the
  penultimate iteration.  Light colors (magenta, cyan): uncorrected.
  Dark colors (red, blue): corrected; the corrected values should be
  zero.\label{F:LCT}}

\end{figure}

\cref{F:LCT} shows examples of leakage scans.  The light markers
indicate the measured (``raw'') current.  With $V_b = 150$~V
(\cref{F:LCT}a), the leakage current is 20 to 30~pA\@.  With $V_b =
-20$~V (\cref{F:LCT}b), the leakage current is smaller in magnitude
and opposite in sign.  The measurements with negative bias are
repeated 120 times, following the same timing algorithm as for the SEY
scans; the current changes by 2 to 3~pA in this time, which is about
2\% of $I_p$ for Phase II parameters.

The dark markers in \cref{F:LCT} show the result of applying the
time-dependent leakage correction.  Over most of the scan, the
corrected current is $\pm 1$~pA or less, which is about 1\% of $I_p$.
There are larger discrepancies during the first few minutes of the
scan, illustrating the usefulness of the iterations.

As can be seen in \cref{F:LCT}b, the time-dependent correction
compensates for the transient behavior reasonably well.  Zooming in on
one iteration (\cref{F:LCT}c), we see that the corrected current
differs from zero, but the systematic differences are comparable to
the noise in the measurement.  The time-dependent correction uses the
recorded time stamp for each current measurement, so that variations
in the time per grid point and timing adjustments to avoid cross-talk
(\autoref{S:cross}) are accounted for.

\begin{figure}[b]
\centering
\includegraphics[width=\widewidth]{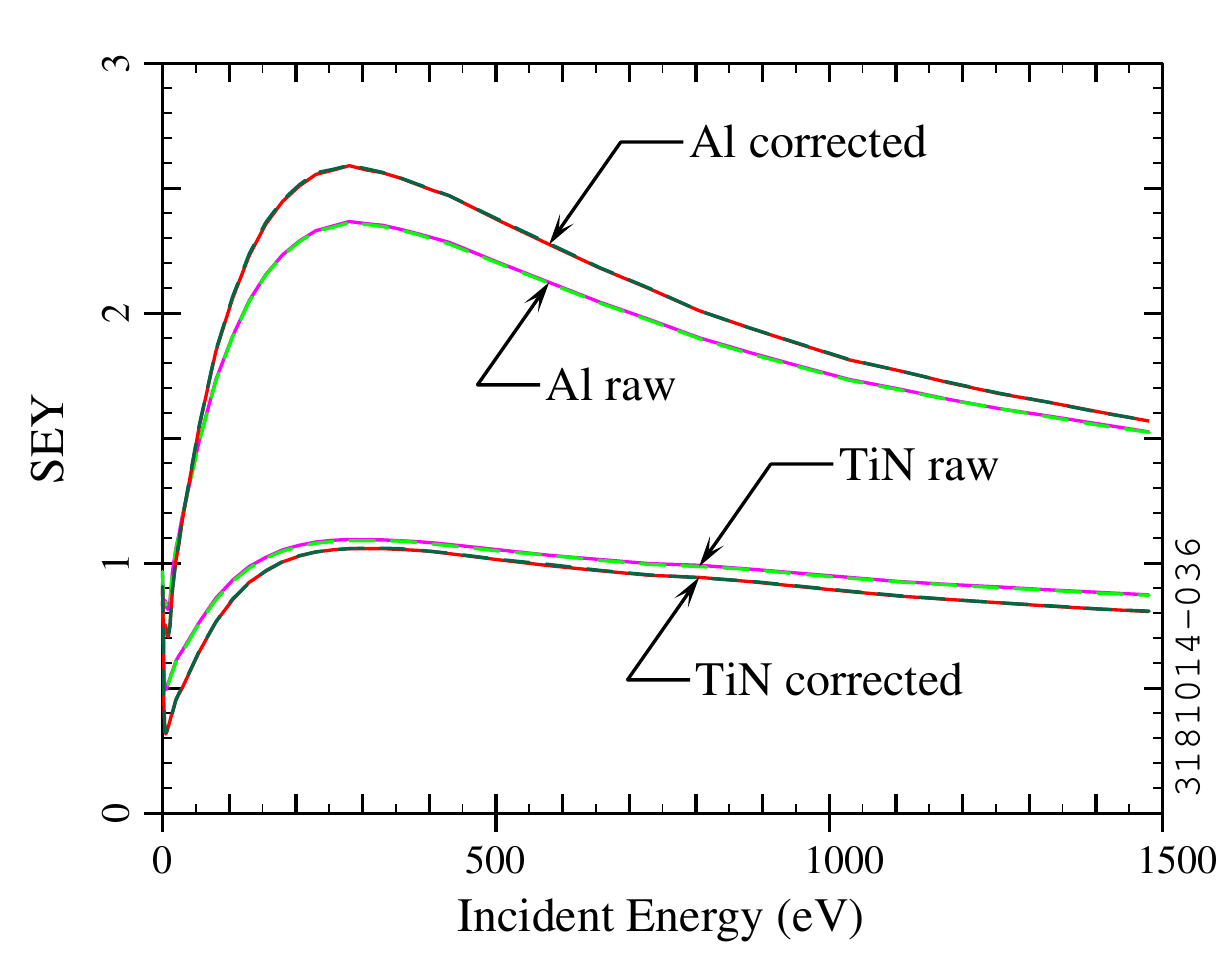}\\[-\bigspace]

\caption[Measurements of SEY as a function of energy without and with
  correction for leakage and transient current]{Measurements of SEY as
  a function of energy without correction (light colors) and with
  correction for leakage current and transient response (dark colors).
  (Middle grid point of 45\degree{} sample, $\theta = 25\degree$.)
  Upper curves: unconditioned Al (Jan 2013).  Lower curves:
  reconditioned TiN (Nov 2012).  Solid curves: first measurement.  Dashed
  curves: repeat measurement.\label{F:SEYcor}}

\end{figure}

Because $I_p$ and $I_t$ are both corrected, the effect on SEY can
partially cancel.  For example, if the uncorrected and corrected $I_t$
values are small relative to $I_p$, SEY $\approx 1$ and the
corrections to $I_p$ produce little change in SEY\@.  \cref{F:SEYcor}
shows examples of the current correction's impact on SEY\@.  For
unconditioned Al with a peak SEY of $\sim 2.5$, the correction
increases the peak by $\sim 10$\%.  For reconditioned TiN with a peak
SEY of $\sim 1$, the correction decreases the peak by $\sim 5$\%.

In \cref{F:SEYcor}, both the first (solid curves) and the repeated
(dashed curves) measurements are shown, since this grid point is
measured twice in the double scan; the first and second $I_t$
measurements are separated in time by $\sim 17$~s.  The $I_t$ values
are corrected by different amounts to account for the current
transient.  However, there is little difference in SEY, which
indicates that the variation in $I_t$ over the time required to scan
the grid points has little impact on SEY\@.

Thus, in the examples above, the magnitude of the leakage current is
$\lappeq 15$\% of $I_p$, which is typical for measurements with
leakage mitigation; the unmitigated leakage current could be $\gappeq
100$\% of $I_p$ under adverse conditions.  The correction in the SEY
due to the leakage current is $\lappeq 10$\%, which is also typical
for Phase II measurements.  With 120 grid points, there is a clear
time dependence in the leakage current due to the bias switch
transient.  The time-dependent leakage correction accounts for this
effectively, but the impact on the SEY is small for Phase II, as we
used a 60~s wait time after a bias switch.  Hence it may be possible
to use a shorter wait time in future measurements.

\subsection{Uncertainties\label{S:uncertain}}

As discussed above, most Phase II modifications were oriented toward
reducing systematic errors.  \autoref{T:ErrSummary} summarizes the
estimated error contributions from various sources for various
scenarios.  The values apply to both $I_p$ and $I_t$, but are
expressed as a percentage of $I_p$ (it is not straightforward to
estimate the error as a fraction of $I_t$).

\begin{table}[b]
\centering
\caption[Estimated current measurement errors]{Summary of estimated
  current measurement errors as a percentage of $I_p$.  For errors due
  to leakage and transient currents, the Phase II value of $I_p \sim
  200$~pA is assumed.  The scenarios used for the final Phase II
  procedure are in bold type.  HH = high humidity, LH = low humidity
  (as quantified in \ref{S:humid}), SC = static correction, TDC =
  time-dependent correction.\label{T:ErrSummary}}
\vspace*{\smallspace}
\begin{tabular}{l|l|l|r}\hline
Source & Mitigate & Correct for & Error \\ \hline\hline
Gun current      & no (Ph. I) & no & $\lappeq 16$\% \\
drift (\S \ref{S:Istab})   & {\bf yes} (Ph. II) & {\bf no} & \boldmath $\lappeq 2\%$ \\ \hline
Leakage          & no (HH) & no & $\gappeq 100$\% \\
current          & no (LH) & no & $\lappeq 14$\% \\
(\S \ref{S:LCmit}--\ref{S:LCmeas})    & no (HH) & yes & $\gappeq 100$\% \\
                 & no (LH) & yes & $\lappeq 1$\% \\
                 & yes & no & $\lappeq 14$\% \\
                 & {\bf yes} & {\bf yes} & \boldmath $\lappeq 1\%$ \\ \hline
Transient        & no ($t_{bw} = 0$) & no & $\gappeq 100$\% \\
current          & yes ($t_{bw} = 60$ s) & no & $\lappeq 4$\% \\
(\S \ref{S:trans}, \S \ref{S:LCC}) & yes ($t_{bw} = 60$ s) & yes (SC) & $\lappeq 2$\% \\
                 & {\bf yes} ($t_{bw} = 60$ s) & {\bf yes} (TDC) & {\boldmath $\lappeq 1\%$} \\ \hline
Cross-talk       & no (Ph. IIa) & no & $\lappeq 50$\% \\
 (\S \ref{S:cross})       & {\bf yes} (Ph. IIb) & {\bf no} & {\bf none} \\ \hline
\end{tabular}

\end{table}

Using \cref{E:seyt}, one can infer the impact of errors in the
measurement of $I_p$ and $I_t$ on the calculated SEY\@.  For Phase
IIb, we expect the items listed in \autoref{T:ErrSummary} to produce a
systematic error in SEY of at most a few percent for $0 \leq$ SEY
$\leq 2$.

Errors due to charging and conditioning are not included in
\autoref{T:ErrSummary}.  From \cref{F:dlc}, we infer that the error
in SEY due to charging was $\sim 45$\% for DLC with the Phase I
method.  With the Phase IIb method, as discussed in
\autoref{S:charge}, we observe some charging or conditioning of
unconditioned and susceptible materials at the parking point, which
decreases the measured SEY by $\lappeq 7$\% with mitigation (see also
\autoref{S:reproduce}).

Overall, we expect the items considered in \autoref{S:phaseii} to
contribute a few percent to the systematic error in the SEY for most
grid points (and $\lappeq 10$\% for the parking point) with the Phase
IIb method.  We estimate that the statistical errors are of the same
order.  A future paper will include more detailed results with a more
complete error analysis.

\section{Examples of SEY Results\label{S:exam}}

Some examples of SEY measurements are presented in this section.  The
beam dose is given in terms of the integrated current of stored $e^-$
bunches; 1 amp$\cdot$hour corresponds to about $3 \cdot
10^{21}$~photons/m of direct SR at the location of the samples.

\subsection{SEY as a Function of Energy}

\begin{figure}[htb]
\centering
\includegraphics[width=\widewidth]{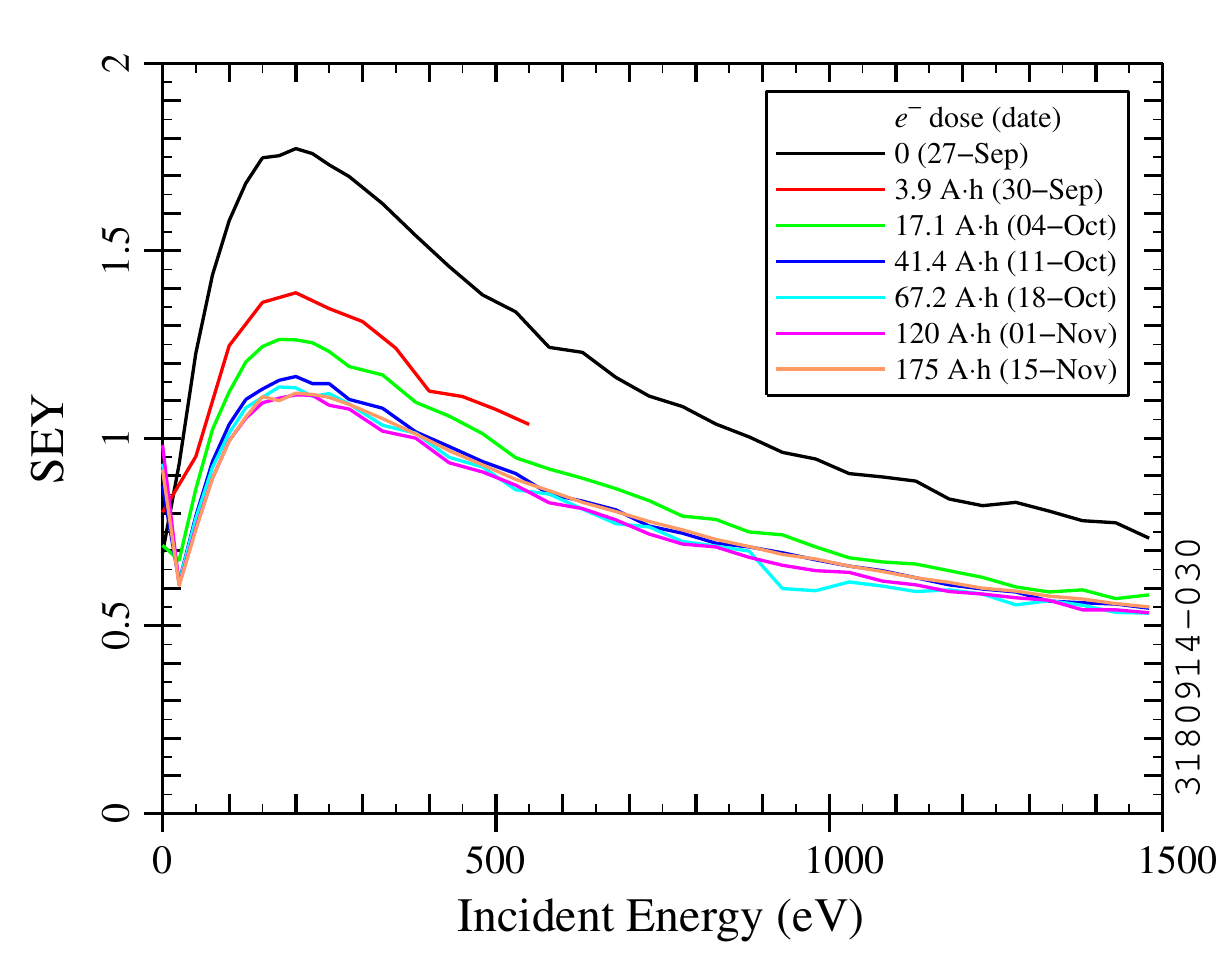}\\[-\bigspace]

\caption[SEY as a function of energy for DLC]{SEY as a function of
  incident energy for the 45\degree{} DLC sample (middle grid point,
  $\theta = 25\degree$, Phase IIa, Sep-Nov 2011).\label{F:SeyEng}}

\end{figure}

\cref{F:SeyEng} shows the measured SEY of the $45\degree$ DLC-coated
sample as a function of energy for different beam doses.  The peak in
the SEY is at about 200~eV\@.  There is a clear decrease in SEY with
beam dose: before conditioning, the peak SEY is about 1.8; for doses
$> 20$~A$\cdot$h, the peak SEY is in the range of 1.1 to 1.2.  The
changes due to conditioning are large compared to the estimated errors
in the measurement.

\subsection{Peak SEY as a Function of Vertical Position}

\begin{figure}[tb]
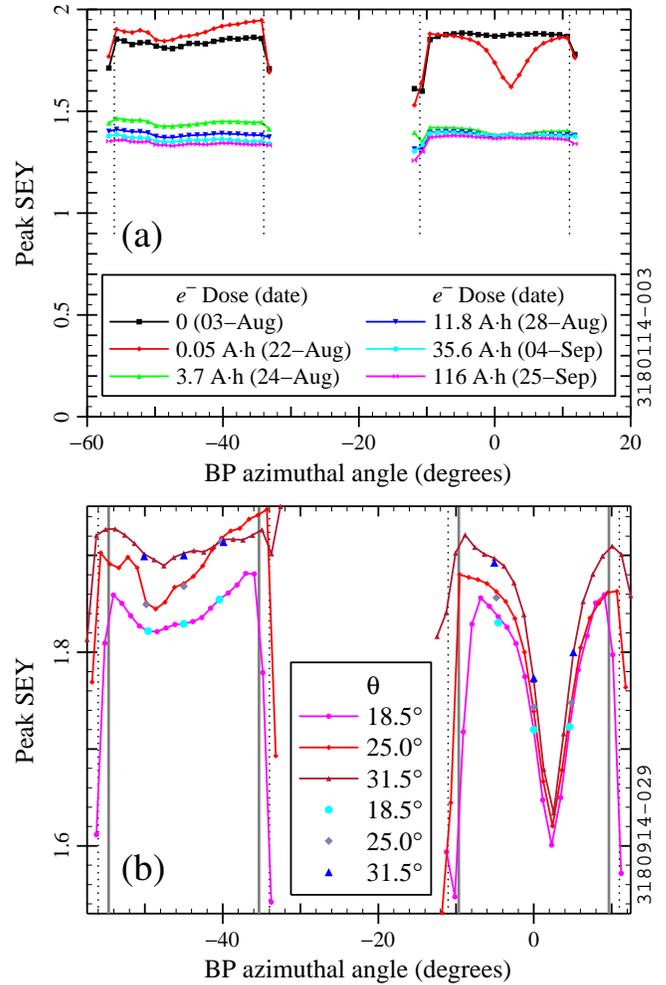

\centering
\GRAFwidth[\middlewidth]{490}{390}
\GRAFoffset{-65}{-65}
\GRAFlabelcoord{25}{125}
\incGRAFlabel{fig14a}{(a)}\\[-3ex]
\GRAFlabelcoord{25}{25}
\incGRAFlabel{fig14b}{(b)}\\[-\bigspace]

\caption[Peak SEY as a function of vertical position for stainless
  steel]{Peak SEY of stainless steel samples (Phase IIb, Aug-Sep 2012)
  as a function of position expressed in terms of the azimuthal angle
  along the beam pipe (BP).  (a) Scans along the middle of the sample
  ($\theta \approx 25\degree$) for different beam doses; (b) scans at
  different incident angles for the 0.05 A$\cdot$h case, with repeated
  points included.  The vertical lines indicate the edges of the
  sample (black dotted: $\theta \approx 25\degree$; solid gray:
  $\theta \approx 18.5\degree, 31.5\degree$).\label{F:SeyHeight}}

\end{figure}

\cref{F:SeyHeight} shows measurements of the peak SEY as a function
of vertical position for stainless steel.  The gun deflection is
converted to azimuthal angle along the beam pipe, with one sample
(horizontal) centered at 0 and the other at $-45\degree$.

\cref{F:SeyHeight}a compares different beam doses.  Before beam
exposure (black), the peak SEY is about 1.8 and is approximately
constant.  After a small beam dose (red), a dip in the SEY appears
near the middle of the horizontal sample, presumably due to direct
SR\@.  For high doses, the SEY decreases and returns to being
approximately independent of position.  The observed differences are
again large compared to the estimated errors.

\cref{F:SeyHeight}b compares the peak SEY as a function of vertical
position for 3 different horizontal deflections (\cref{F:GridSamXY},
square markers) for the 0.05 A$\cdot$h case (the red case in
\cref{F:SeyHeight}a).  The legend indicates the approximate angle of
incidence ($\theta$).  The dip in peak SEY at $\sim 2\degree$ is seen
in all 3 scans.

\subsection{Reproducibility\label{S:reproduce}}

Nine grid points are measured twice (as seen in \cref{F:GridSamXY}),
which provides a check of the short-term reproducibility of the
measurements.  In \cref{F:SeyHeight}b, vertical scans
(\cref{F:GridSamXY}, squares) are shown in red; repeated points from
horizontal scans (\cref{F:GridSamXY}, diamonds) are shown in blue.
The repeated values are reasonably consistent.  This indicates that
the features seen in the vertical scans are reproducible and that the
system is able to properly resolve the dependence of SEY on position.

We can repeat measurements occasionally during extended access
periods.  Repeated measurements on Al samples (1 or 2 days apart,
without intervening exposure to beam) in Phase II indicate that the
measured SEY can vary by up to 5\% or more for a few grid points.  For
most of the 120 grid points, the SEY varies by a few percent or less,
consistent with what we expect based on the systematic uncertainties
(\autoref{S:uncertain}).  Thus the measured changes from scrubbing are
large compared to the day-to-day reproducibility of the measurements.

\section{Conclusion}

We have developed an in-situ secondary electron yield measurement
system to observe conditioning of metal and coated samples by CESR
beams.  We have made iterative improvements in the measurement method
to reduce charging and conditioning by the electron gun; mitigate and
correct for the leakage current and transient current; eliminate
cross-talk between the adjacent SEY stations; and mitigate gun current
drift.  We have reduced the systematic error due to these effects to a
few percent, allowing us to measure the dependence of the SEY on beam
dose, position, and angle with better resolution.

There is room for additional improvement in the techniques.  Sources
of systematic error that we have not yet accounted for include (i) the
escape of elastic and rediffused secondaries during the measurement of
the primary current, and (ii) the deflection of low-energy primary
electrons by the sample bias.  A more direct measurement method may
help further reduce the systematic errors.  Measuring the energy
distribution of the secondary electrons would allow us to distinguish
between elastic, rediffused, and true secondaries.  Additional
improvements to the apparatus and techniques might allow us to reduce
the measurement time and decrease the incidence of noise spikes in the
current due to nearby activity.  Some of the improvements described
above may not be practical for our in-situ apparatus, and may require
an out-of-tunnel SEY measurement system.

Our ultimate goal is to use the SEY measurements under realistic
conditions to constrain the SEY model parameters as much as possible;
this will help improve the predictive ability of models for electron
cloud build-up, allowing for more successful electron cloud mitigation
in future accelerators, so that they can achieve better performance
and higher reliability.

\section*{Acknowledgments}
\addcontentsline{toc}{section}{Acknowledgments}

We are grateful for the support of collaborators at SLAC, who provided
hardware, samples, and guidance for the SEY studies at \CesrTA.
Coatings of samples with aC and DLC were done by CERN and KEK,
respectively.  We thank our collaborators at Fermilab for useful
discussions.

Our work would not have been possible without support from the design,
electronics, fabrication, information technology, operations, survey,
technical services, and vacuum groups.  We are particularly thankful
for the work by V. Medjidzade and the help from W. J. Edwards,
B. M. Johnson, J. A. Lanzoni, R. Morey, and R. J. Sholtys.  We thank
our \CesrTA{} collaborators for their support and ideas, particularly
J. R. Calvey, J. A. Crittenden, G. F. Dugan, J. P. Sikora, and
K. G. Sonnad.  S. T. Wang provided valuable help and guidance with our
data acquisition program development work.  We thank S. B. Foster for
doing off-line SEY measurements and helping with in-situ measurements.
We appreciate the management support for our studies, particularly
from M. G. Billing, D. H. Rice, D. L. Rubin, J. W. Sexton, and
K. W. Smolenski.


This work was supported by the National Science Foundation through
Grants PHY-0734867 and PHY-1002467 and by the Department of Energy
through Grants DE-FC02-08ER-41538 and
DE-SC0006505.

\appendix
\section{Measurements and Modifications Time-Line\label{S:hist}}

\autoref{T:hist} outlines the time-line for SEY measurements and
system modifications in Phase I and Phase II.

\begin{table}[htb]
\centering
\caption[SEY measurements and improvements in techniques]{SEY
  in-situ measurements and improvements in techniques.  GCD = gun 
  current drift; resol.\ = resolution; R \& R = resolution and
  range.\label{T:hist}}
\vspace*{\smallspace}
\begin{tabular}{lll} \hline
Samples & Dates & Comments \\ \hline\hline
\multicolumn{3}{r}{\bf Phase I Measurements} \\ \hline
TiN & Jan-Aug 2010 & Commission systems \\ \hline
Al 6061 & Aug-Nov 2010 & Remove vacuum gauges \\ \hline
aC & Nov 2010- & \\
 & Jan 2011 & \\ \hline
\multicolumn{3}{r}{\bf Phase IIa Development} \\ \hline
(none) & Jan-Aug 2011 & Ensure direct SR (\S \ref{S:protrude}); mi- \\
 &  & tigate GCD (\S \ref{S:Istab}), charging\\
 & & (\S \ref{S:charge}), \& leakage (\S \ref{S:LCmit}--\ref{S:trans}) \\ \hline
\multicolumn{3}{r}{\bf Phase IIa Measurements} \\ \hline
DLC & Sep-Nov 2011 & Investigate spatial resol.\ \\ \hline
TiN & Nov 2011- & Improve spatial resol.\ (\S \ref{S:hshake}); \\
2nd pair & Mar 2012 & variable energy step (\S \ref{S:EngRes});  \\
 & & investigate spatial R \& R \\ \hline
Cu & Mar-Jul 2012 & Improve spatial R \& R \\ \hline
\multicolumn{3}{r}{\bf Phase IIb Measurements} \\ \hline
Stainless & Aug-Sep 2012 & Full spatial R \& R (\S \ref{S:pts}); mi- \\
steel &  & tigate parasitic conditioning\\
316 &  &  (\S \ref{S:charge}) \& cross-talk (\S \ref{S:cross})\\ \hline
TiN & Oct 2012- & Recondition after air \\
2nd pair & Jan 2013 & exposure \\ \hline
Al 6063 & Jan 2013- & Long-term conditioning \\ \hline
\end{tabular}
\end{table}

\section{Model for Leakage and Transient Current\label{S:derive}}

The leakage current changes slowly in the time required to measure
$I_t$ for all of the grid points (\autoref{S:LCC}).  This led us to
develop a model to account for both the leakage current and the
transient current.  Additional details are provided in
\cite{ARXIV:1407.0772}.

We measured the sample current $I$ as a function of time $t$ after
stepping the bias voltage.  For a circuit with resistive and
capacitive elements, the current should decay exponentially towards a
steady-state value $I_\infty$ after a bias step.  This was not a good
description of the current we measured.  Our inference is that the
picoammeter is an active element of the circuit.

After a step from $V_b = V_0$ to $V_b = V_2$ at time $t_1$, we found
\begin{equation}
I(t) \approx \Gamma_\parallel \left(\frac{V_2 - V_0}{t - t_1}\right) + \frac{V_2}{R_\parallel} \; ,\label{E:ICeRp}
\end{equation}
where $\Gamma_\parallel$ and $R_\parallel$ are constants.  For $t
\rightarrow \infty$, $I(t) \rightarrow I_\infty = V_2/R_\parallel$,
consistent with a simple circuit with resistance $R_\parallel$ from
sample to ground.  The constant $\Gamma_\parallel$ has dimensions of
capacitance, and can be thought of as being a ``capacitance-like''
quantity.  Since \cref{E:ICeRp} is not derived from a circuit, we
refer to it as a ``semi-empirical'' model.

\begin{figure*}
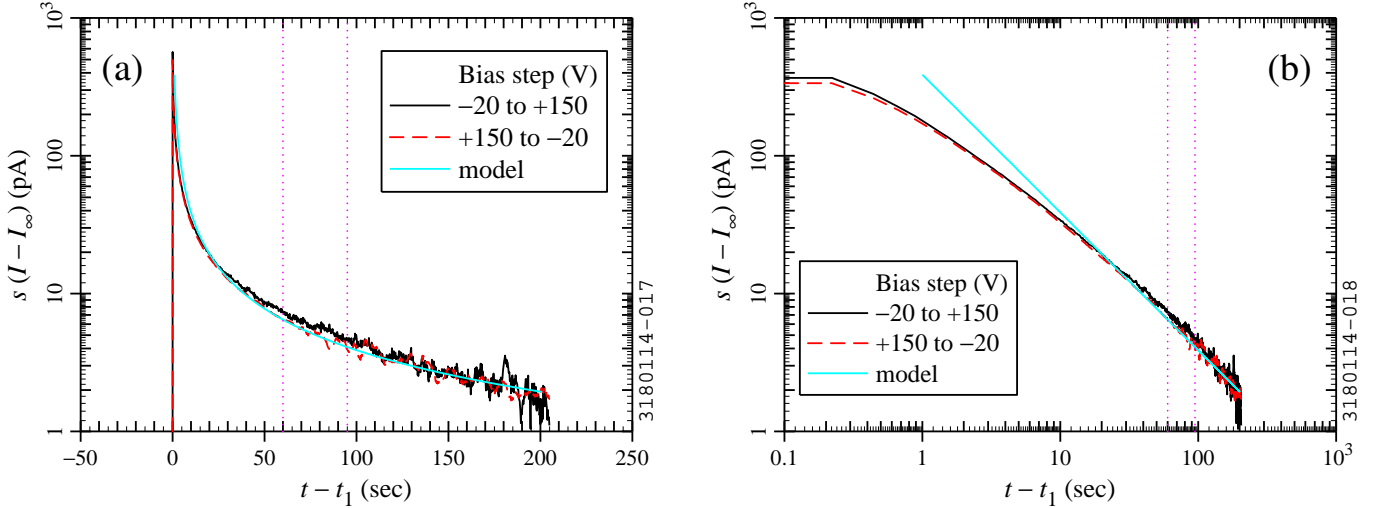

\centering
\GRAFwidth[0.49\textwidth]{490}{390}
\GRAFoffset{-65}{-65}
\GRAFlabelcoord{15}{255}
\incGRAFlabel{fig15a}{(a)}%
\hspace*{\fill}
\GRAFlabelcoord{350}{255}
\incGRAFlabel{fig15b}{(b)} \\[-\bigspace]

\caption[Measured current as a function of time with a bias
  step]{Measured current as a function of time for the 45\degree{} SEY
  station with a step in the sample bias between $-20$~V and $+150$~V
  at time $t_1$, with (a) linear-log, and (b) log-log scale.  The
  solid cyan curve represents the model, using best match parameters
  of $\Gamma_\parallel = 2.3$~pF and $R_\parallel =
  10$~T$\Omega$.\label{F:dIvtreal}}

\end{figure*}

\cref{F:dIvtreal} shows measurements on the 45\degree{} system.
The time at which the bias is stepped ($t_1$) is subtracted from $t$
and the steady-state current ($I_\infty$) is subtracted from $I(t)$.
Additionally, $I - I_\infty$ is multiplied by a sign correction
coefficient $s = \pm 1$ to compare upward and downward steps in $V_b$
on the same footing.

It is clear from the log-linear scale of \cref{F:dIvtreal}a that
the current does not decrease exponentially to its steady-state value.
The transient current is as high as about $\pm 500$~pA, which is much
larger than the steady-state current of $\pm 15$~pA or less.

\cref{F:dIvtreal}b shows the current as a function of time on a
log-log scale.  The cyan curves represent the semi-empirical model,
which fits the measurements reasonably well when $t - t_1 > 30$~s.
For $t - t_1 < 30$~s, the model differs significantly from the
measured current.  However, the relevant time interval for SEY
measurements starts 60~s after the bias step and lasts for 35~s
(indicated by the dotted lines).  Hence the discrepancies for $t - t_1
< 30$~s are not a problem for our Phase II procedure.

The parameter values to fit the measured current are in the range of
$\Gamma_\parallel = 0.5$ to $3$~pF and $R_\parallel = 6$ to
$25$~T$\Omega$; per \ref{S:LCtrend}, the best-fit values are different
for each SEY station and vary with time, leading us to do a leakage
scan prior to each SEY scan (\autoref{S:LCmeas}).  We use the
parameter values from the leakage scan to correct the currents for the
subsequent SEY scan, as described in more detail in
\cite{ARXIV:1407.0772}.

\section{Unmitigated Leakage Current\label{S:humid}}

In Phase IIb, some measurements without leakage mitigation were done
to better understand the final systems' behavior.  We measured the
leakage on the off-line station at various humidities (set by the
outside air conditions as modified by the climate control system) with
the N$_2$ gas flow off; we measured the relative humidity (RH) using a
portable hygrometer.\footnote{Model 4189, Control Company,
  Friendswood, TX.}

\begin{figure}[b]
\centering
\includegraphics[width=\widewidth]{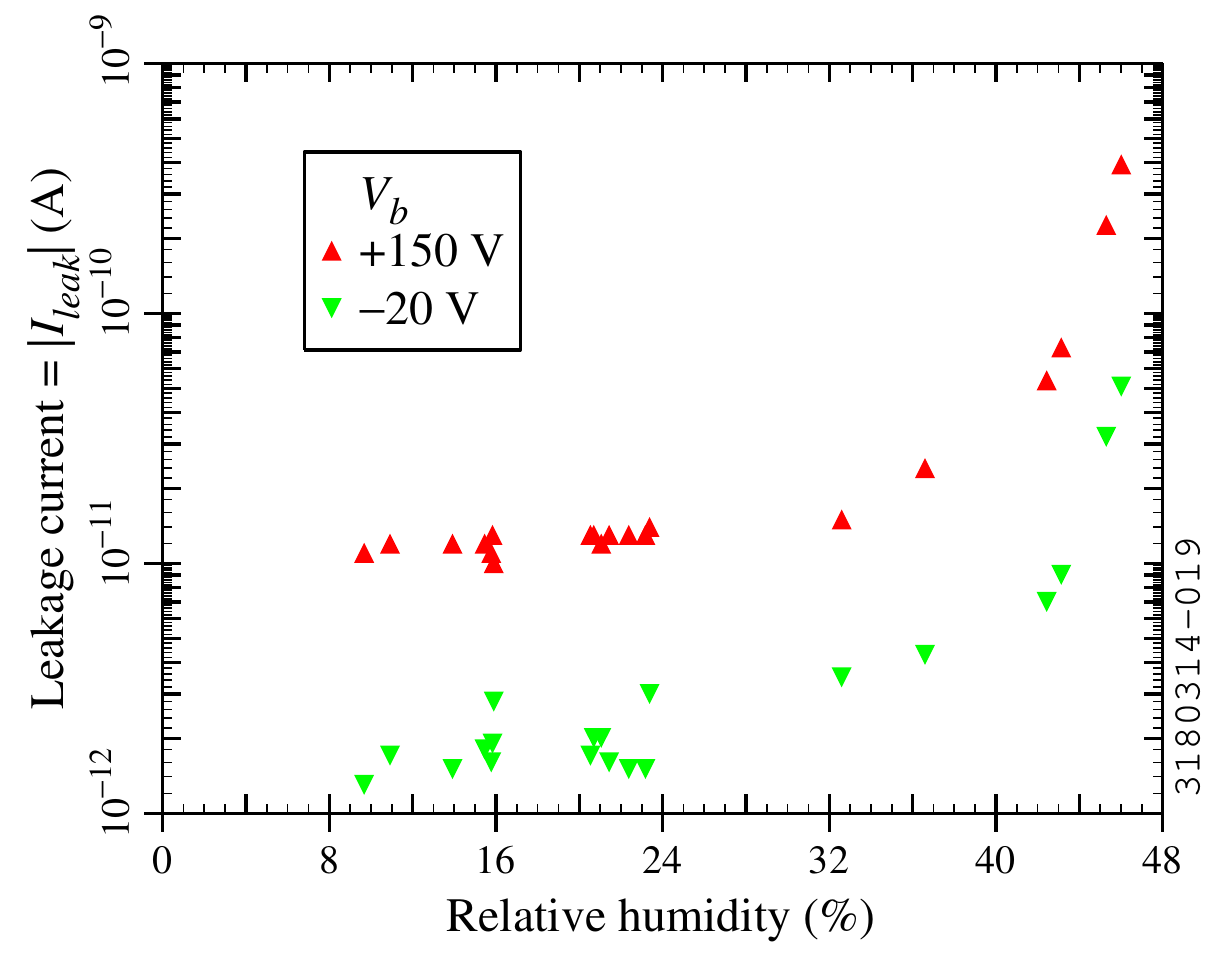}\\[-\bigspace]

\caption[Measured leakage current as a function of humidity without
  mitigation]{Measured leakage current as a function of humidity for
  the off-line station without mitigation (temperature = 21.5 to
  23.5\degree C).\label{F:LChumid}}

\end{figure}

\cref{F:LChumid} shows the leakage current ($I_{leak}$) as a
function of humidity for $V_b = 150$~V and $-20$~V\@.  For RH $\gappeq
30$\%, the leakage increases rapidly, changing by more than a factor
of 10 between the lowest and highest humidities.  For RH $\lappeq
30$\%, the leakage is low, though it still shows some variation.

We can infer the resistance to ground ($R_{leak}$) from the bias and
the leakage current.  The calculated values of $R_{leak}$ for positive
and negative bias are roughly consistent; per \ref{S:derive}, ideally
we should have $R_{leak} = R_\parallel$.  At low humidity, $R_{leak}$
is in the range of 10 to 20 T$\Omega$, consistent with the values of
$R_\parallel$ with mitigation of \ref{S:LCtrend}.

\begin{figure}
\centering
\includegraphics[width=\widewidth]{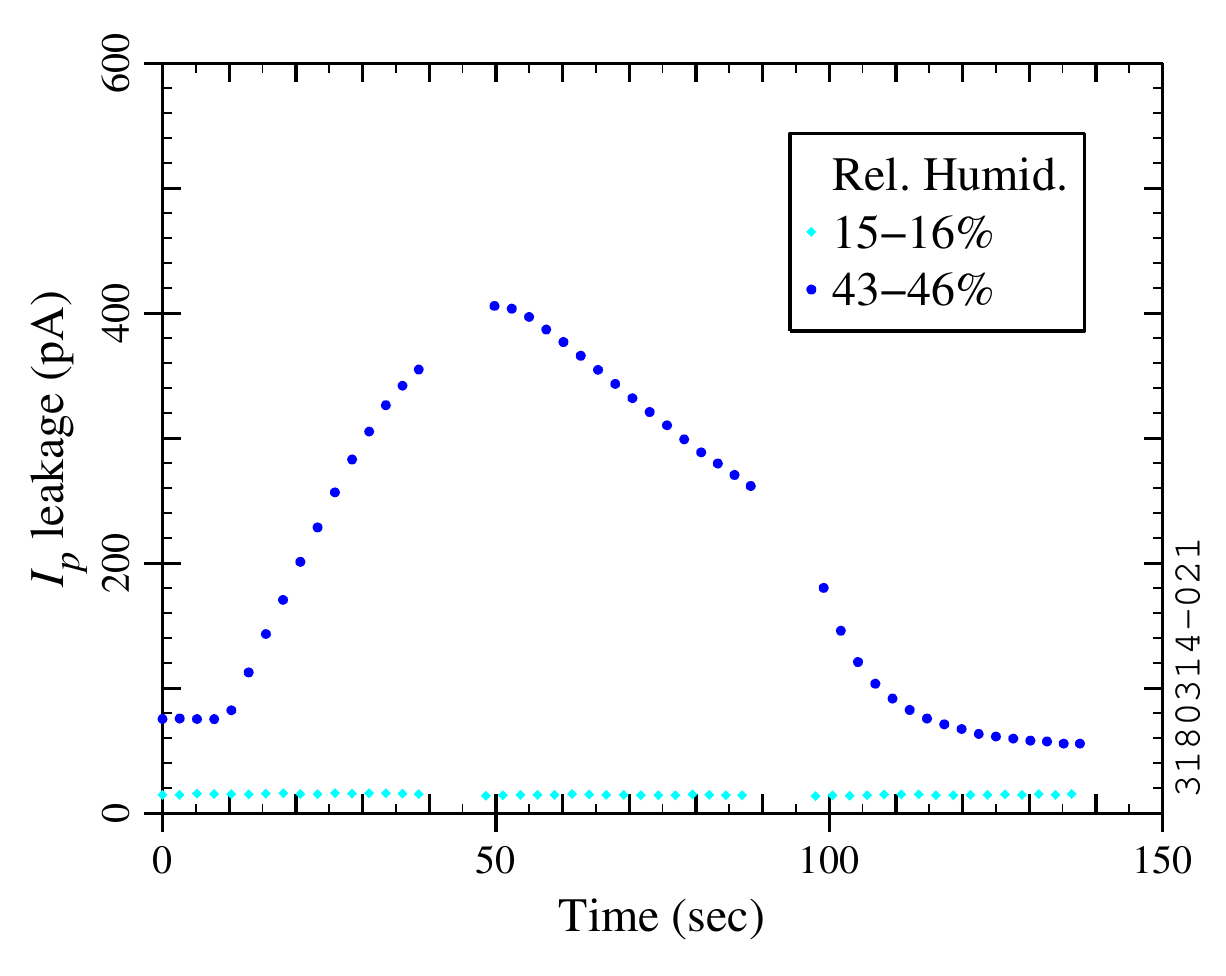}\\[-\bigspace]

\caption[Leakage scans for the off-line station without mitigation at
  different ambient humidities]{Leakage scans for the off-line station
  without mitigation at different ambient humidities ($V_b = +150$~V,
  temperature = 21.5 to 23.5\degree C).\label{F:LClowmedH}}

\end{figure}

For RH $\gappeq 35$\%, \cref{F:LChumid} shows that a small change in
humidity produces a large change in leakage.  Hence SEY measurements
in a humid environment without a gas blanket are likely to have large
errors due to small humidity variations.  \cref{F:LClowmedH} shows
leakage scans without gas flow at different humidities.  With low
humidity (light blue), RH decreased from 15.9\% to 15.5\% during the
scan.  The leakage remained low and stable.  With high humidity (dark
blue), RH varied from 43.1\% up to 46.0\% and then down to 42.5\%.
Correspondingly, the leakage varied by a factor of $\sim 4$.
Measurements at $V_b = -20$~V (not shown) were interleaved with the
measurements at $+150$~V, with similar trends.  Thus, as anticipated,
a humid environment gives poor leakage current stability.

We conclude that, in a dry environment (RH $\lappeq 30$\%), the
leakage current for our SEY stations is relatively stable, and
mitigation is not needed.  With higher humidity and no mitigation,
leakage corrections are large: at RH = 46\%, the leakage with $V_b =
+150$~V exceeds our Phase II value of $I_p \approx 200$~pA, and varies
significantly over the time of an SEY scan.  A dry nitrogen blanket
ensures low and stable leakage current.  However, a dry environment
does not remove the need to measure the leakage current; even with the
N$_2$ blanket, we observe long-term variation in the leakage, as
discussed in the next section.

\section{Mitigated Leakage Current\label{S:LCtrend}}

\begin{figure}
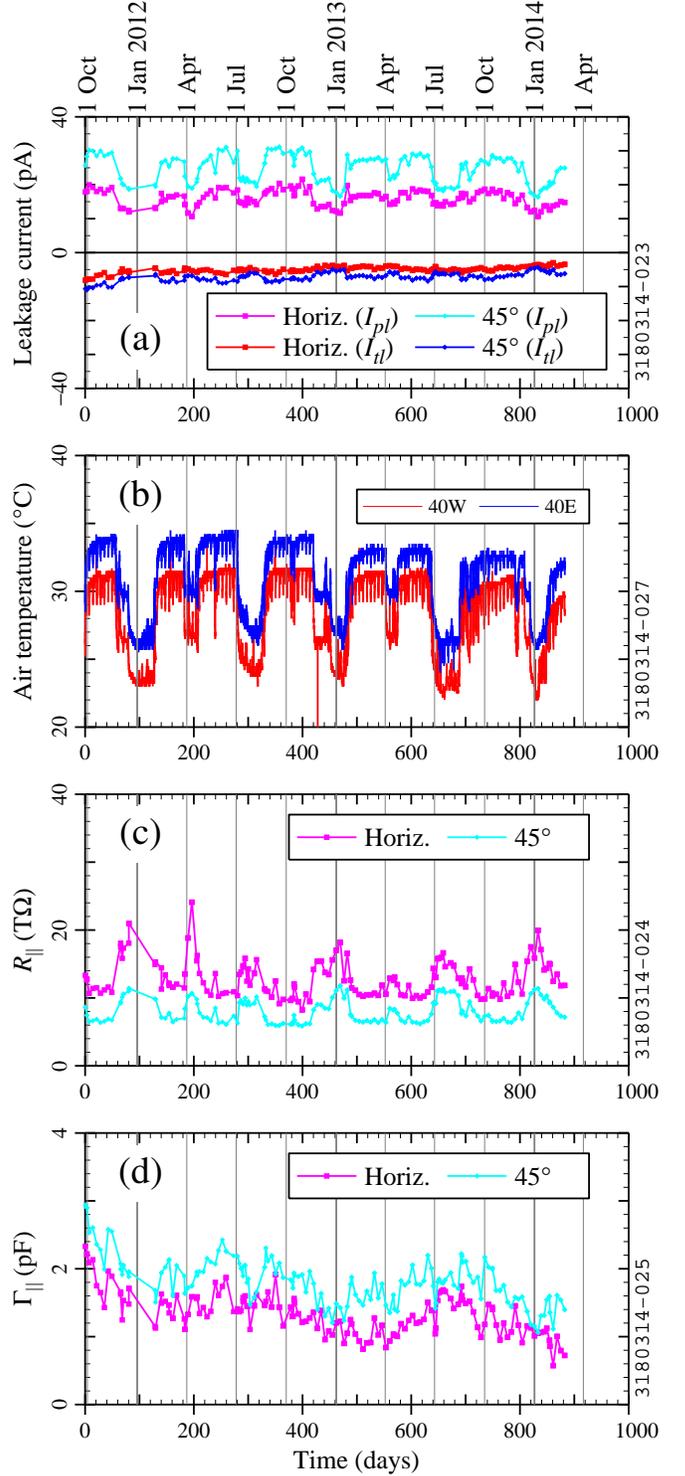

\centering
\GRAFwidth[\narrowwidth]{490}{313}
\GRAFoffset{-65}{-25}
\GRAFlabelcoord{15}{20}
\GRAFlabelbox{40}{30}
\incGRAFboxlabel{fig18a}{(a)}\\[-0.3ex]
\GRAFwidth[\narrowwidth]{490}{250}
\GRAFlabelcoord{15}{155}
\incGRAFboxlabel{fig18b}{(b)}\\[-0.3ex]
\incGRAFboxlabel{fig18c}{(c)}\\[-0.3ex]
\GRAFwidth[\narrowwidth]{490}{290}
\GRAFoffset{-65}{-65}
\incGRAFboxlabel{fig18d}{(d)}\\[-\medspace]

\caption[Comparison of long-term leakage trends]{Comparison of
  long-term trends: (a) measured leakage currents, (b) tunnel air
  temperature, (c, d) leakage model parameters as a function of time.
  The measurements were done over approximately 2.5 years, from 27 Sep
  2011 (time = 0) to 25 Feb 2014 (time = 882 days).  The gray lines
  correspond to quarterly calendar dates.  Temperatures were measured
  with thermocouples in the tunnel at 40E ($\sim 10$~m East of the SEY
  stations) and 40W ($\sim 10$~m to the West of the
  stations).\label{F:LCtrend}}

\end{figure}

In Phase II, we used an N$_2$ gas blanket to mitigate leakage
(\autoref{S:LCmit}) and did a leakage scan prior to each SEY scan
(\autoref{S:LCmeas}).  Measured leakage currents are shown in
\cref{F:LCtrend}a: $I_{pl}$ and $I_{tl}$ were measured with $V_b =
+150$~V and $-20$~V, respectively.  As can be seen, the leakage
current has varied by a factor of $\sim 2$ in Phase II\@.  The leakage
current changes enough over time to make repeated leakage scans
necessary---a constant-leakage assumption would introduce significant
systematic errors into the SEY calculation.

The leakage current varies non-randomly, though it does not decrease
steadily or change seasonally.  \cref{F:LCtrend}a suggests that the
leakage current is higher when the accelerator is running and lower
during summer and winter down periods.  One difference between
high-current operation and down periods is the tunnel air temperature,
as illustrated in \cref{F:LCtrend}b.  The air temperature increases by
about 8\degree C when CHESS currents are stored.  Comparison of
Figs.~\ref{F:LCtrend}a and \ref{F:LCtrend}b shows that the
leakage current is indeed correlated with temperature.  This
correlation could come about if the leakage properties of the ceramic
or stand-offs are temperature-dependent, or if the moisture content of
the gas blanket is temperature-sensitive.

Figs.~\ref{F:LCtrend}c and \ref{F:LCtrend}d show the model parameters
(see \ref{S:derive}) calculated from the measured leakage currents.
The resistance to ground ($R_\parallel$) shows a clear inverse
correlation with the measured leakage, as expected; $\Gamma_\parallel$
also shows a time dependence, though it is more difficult to
interpret.  \cref{F:LCtrend}d suggests that $\Gamma_\parallel$ has
some seasonal correlation and a slight downward trend, which might be
artifacts of the semi-empirical nature of the model for the transient
response.



\bibliographystyle{compact}
\bibliography{BibListAll}


\end{document}